\documentclass[PAPER, cernpreprint, USenglish, texmf, orcidlogo]{atlasdoc}

\usepackage[subcaption]{atlaspackage}
\usepackage{atlasbiblatex}
\usepackage{acronym}   %
\usepackage{multirow}  %
\usepackage[hepparticle,hepprocess]{atlasphysics}

\graphicspath{{logos/}{figures/}}

\usepackage{ANA-TOPQ-2021-26-PAPER-defs}
\usepackage[capitalise]{cleveref}

\AtlasTitle{Measurement of top-quark pair production in association with charm quarks in proton--proton collisions at $\sqrt{s}=13$ TeV with the ATLAS detector}

\AtlasAbstract{%
Inclusive cross-sections for top-quark pair production in association with charm quarks are measured with proton--proton collision data at a center-of-mass energy of 13\,TeV corresponding to an integrated luminosity of 140\,fb$^{-1}$, collected with the ATLAS experiment at the LHC between 2015 and 2018.
The measurements are performed by requiring one or two charged leptons (electrons and muons), two $b$-tagged jets, and at least one additional jet in the final state.
A custom flavor-tagging algorithm is employed for the simultaneous identification of $b$-jets and $c$-jets.
In a fiducial phase space that replicates the acceptance of the ATLAS detector, the cross-sections for $t\bar{t}+ {\geq} 2c$ and $t\bar{t}+1c$ production are measured to be
$1.28^{+0.27}_{-0.24}\;\text{pb}$ and
$6.4^{+1.0}_{-0.9}\;\text{pb}$, respectively.
The measurements are primarily limited by uncertainties in the modeling of inclusive $t\bar{t}$ and $t\bar{t}+b\bar{b}$ production, in the calibration of the flavor-tagging algorithm, and by data statistics.
Cross-section predictions from various $t\bar{t}$ simulations are largely consistent with the measured cross-section values, though all underpredict the observed values by 0.5 to 2.0 standard deviations.
In a phase-space volume without requirements on the $t\bar{t}$ decay products and the jet multiplicity, the cross-section ratios of $t\bar{t}+ {\geq} 2c$ and $t\bar{t}+1c$ to total $t\bar{t}+\text{jets}$ production are determined to be $(1.23 \pm 0.25) \%$ and $(8.8 \pm 1.3) \%$.
}

\author{The ATLAS Collaboration}

\AtlasRefCode{TOPQ-2021-26}

\PreprintIdNumber{CERN-EP-2024-242}

\arXivId{2409.11305}

\HepDataRecord{https://www.hepdata.net/record/ins2829504}

\AtlasJournal{\PLB}
\AtlasJournalRef{\PLB 860 (2025) 139177}
\AtlasDOI{10.1016/j.physletb.2024.139177}

\hypersetup{pdftitle={ATLAS document},pdfauthor={The ATLAS Collaboration}}

\begin{document}

\maketitle

\section{Introduction}

\begin{figure}
\centering
\subcaptionbox{\label{fig:ttcc_diagrams_a}}{\includegraphics[width=0.30\textwidth]{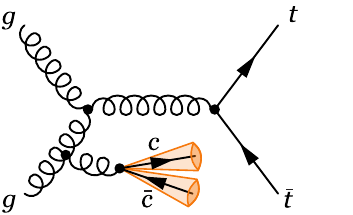}}
\subcaptionbox{\label{fig:ttcc_diagrams_b}}{\includegraphics[width=0.30\textwidth]{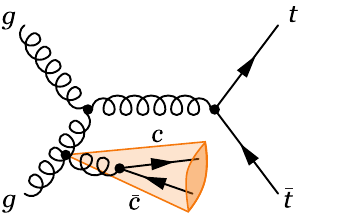}}
\subcaptionbox{\label{fig:ttcc_diagrams_c}}{\includegraphics[width=0.30\textwidth]{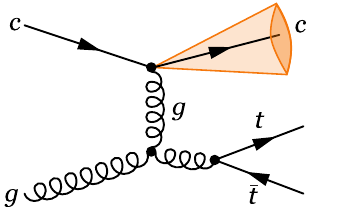}}
\caption{Illustrative Feynman diagrams for \ttccin and \ttc production: \protect\subref{fig:ttcc_diagrams_a} \ttcc production via initial-state gluon radiation where both \cquarks form a jet each, \protect\subref{fig:ttcc_diagrams_b} \ttcc production via initial-state gluon radiation where the two \cquarks are in the same jet, \protect\subref{fig:ttcc_diagrams_c} \ttc production where the \cquark originates from the initial state.}
\label{fig:ttcc_diagrams}
\end{figure}

Due to its pivotal role in the \ac{SM} of particle physics and its potential interactions with new physics in various beyond-\ac{SM} frameworks,
the study of the top quark~($t$) remains a cornerstone of contemporary particle physics research.
In particular, the ATLAS~\cite{PERF-2007-01} and CMS~\cite{CMS-CMS-00-001} experiments at the \ac{LHC} have directed their attention towards exploring rare final states involving top-quark pairs (\ttbar), such as associated production of \ttbar with Higgs bosons (\ttH) or gauge bosons (\ttW, \ttZ, \ttgamma), and four-top-quark production (\tttt).
Measurements of \ttH with $H\to\bbbar$ decay and \tttt topologies with single-lepton or dilepton final states~\cite{CMS-HIG-17-022,CMS-HIG-17-026,HIGG-2020-24,TOPQ-2020-10,CMS-TOP-21-005} encounter significant challenges due to substantial, irreducible background contributions stemming from \ttbar production in association with heavy-flavor quarks, namely bottom ($b$) and charm ($c$) quarks.
These heavy-flavor quarks can be produced via radiation of a gluon which then splits into a \bbbar or \ccbar pair.
Depending on the kinematics, the \bbbar or \ccbar pair can form a jet each ($\ttbb/\ccbar$) or merge into a single jet ($\ttB/1C$).
Single \bquark or \cquark production ($\ttb/c$) can also occur via a \bquark or \cquark originating from the initial state.
Illustrative Feynman diagrams for these three production modes are shown in \cref{fig:ttcc_diagrams}.

Measurements of \ttbb and \ttcc production have been limited by the challenging modeling of these processes.
While computations of the \ttbb production \xsec are available at \ac{NLO} in \ac{QCD}~\cite{Bredenstein:2008zb,Bredenstein:2009aj,Cascioli:2013era,Jezo:2018yaf,Buccioni:2019plc,Bevilacqua:2022twl}, the uncertainties in the choice of the renormalization and factorization scales, denoted by \muR and \muF, remain sizable, primarily due to the distinct energy scales associated with the \ttbar pair and the \bbbar pair.
Experimental measurements of \ttbb production were conducted in proton--proton (\pp) collision data at the \ac{LHC} by the ATLAS and CMS experiments at a center-of-mass energy of $\sqs = \SI{13}{\TeV}$~\cite{TOPQ-2017-12,CMS-TOP-18-011,CMS-TOP-22-009-upd,TOPQ-2019-03}.
These measurements, along with those of \ttH ($H\to\bbbar$) and \tttt production, frequently involve determining the \ttcin normalization \emph{in situ} through a free parameter in the fit, compensating for the limited knowledge of this process.
Recent ATLAS measurements of \ttbb and \ttH($H\to\bbbar$)~\cite{TOPQ-2019-03,HIGG-2020-24} reported the \ttcin normalization factor to be larger than the value predicted by \ac{MC} simulations.
CMS performed a dedicated \ttcc measurement in \ttbar final states with two charged leptons based on \RunTwo data corresponding to \SI{41.5}{\ifb}~\cite{CMS-TOP-20-003}, but did not explicitly measure the \ttcC \xsec.
The rates of \ttbb, \ttcc, and \ttlight~jets were found to agree with predictions from \ac{MC} simulations performed at \ac{NLO} in the matrix element interfaced with a \ac{PS} algorithm (\acused{NLO+PS}\acs{NLO+PS}) within one to two standard deviations of measurement uncertainties.

Using the complete \ac{LHC} \RunTwo data sample of \pp collisions corresponding to an integrated luminosity of \SI{140}{\ifb}, this Letter presents the first ATLAS measurement of \ttcin production.
The production rates of \ttccin and \ttc are determined separately to facilitate more detailed comparisons with theoretical predictions, as these processes are expected to be sensitive to different production mechanisms (c.f.\ \cref{fig:ttcc_diagrams}).
The first comprises all final states with two or more \cjets, e.g., \ttcc production where both \cquarks form separate jets, while the second includes both \ttc production via a \cquark from the initial state and \ttC production.
Events with one or two charged leptons in the final state are considered, targeting  \ttbar topologies where one or both \Wboson~bosons from the top quarks decay leptonically.
While topologies with one charged lepton (\slepton) offer larger statistics, they introduce additional complexity due to the potential production of extra \cquarks from the hadronically decaying \Wboson~boson.
Conversely, final states with two charged leptons (\dilepton) are rarer but expect less background contamination.

The probabilities for \ttbar pairs with additional jets initiated by \bquarks, \cquarks, light quarks, and gluons to enter each analysis region are estimated through \ac{NLO+PS} simulations of inclusive \ttbar and of \ttbb production.
This measurement uses a custom flavor tagging algorithm, termed the \bctagger, tailored to simultaneous \cjet and \bjet identification, to define analysis regions sensitive to \ttccin and \ttc production.
The rates of these processes are then measured in a fiducial phase space designed to replicate the acceptance of the ATLAS detector and in a more inclusive volume.
Furthermore, the ratios of \ttccin, \ttc, and \ttbin to overall \ttjets production are extracted and compared with \ac{NLO+PS} simulations.

\section{ATLAS detector}

The ATLAS experiment~\cite{PERF-2007-01} at the \ac{LHC} is a multipurpose particle detector with a forward--backward symmetric cylindrical geometry and a near \(4\pi\) coverage in solid angle.\footnote{%
ATLAS uses a right-handed coordinate system with its origin at the nominal \ac{IP} in the center of the detector and the \(z\)-axis along the beam pipe.
The \(x\)-axis points from the \ac{IP} to the center of the \ac{LHC} ring, and the \(y\)-axis points upwards.
Polar coordinates \((r,\phi)\) are used in the transverse plane, \(\phi\) being the azimuthal angle around the \(z\)-axis.
The pseudorapidity is defined in terms of the polar angle \(\theta\) as \(\eta = -\ln \tan(\theta/2)\) and is equal to the rapidity
$ y = \frac{1}{2} \ln \left( \frac{E + p_z}{E - p_z} \right) $ in the relativistic limit.
Angular distance is measured in units of \(\Delta R \equiv \sqrt{(\Delta y)^{2} + (\Delta\phi)^{2}}\).
}
It consists of an \ac{ID} surrounded by a thin superconducting solenoid providing a \qty{2}{\tesla} axial magnetic field, electromagnetic and hadronic calorimeters, and a muon spectrometer.
The \ac{ID} covers the pseudorapidity range \(|\eta| < 2.5\).
It consists of silicon pixel, silicon microstrip, and transition radiation tracking detectors.
Lead/liquid-argon (LAr) sampling calorimeters provide \ac{EM} energy measurements with high granularity within the region \(|\eta|< 3.2\).
A steel/scintillator-tile hadronic calorimeter covers the central pseudorapidity range (\(|\eta| < 1.7\)).
The endcap and forward regions are instrumented with LAr calorimeters for \ac{EM} and hadronic energy measurements up to \(|\eta| = 4.9\).
The muon spectrometer surrounds the calorimeters and is based on three large superconducting air-core toroidal magnets with eight coils each.
The field integral of the toroids ranges between \num{2.0} and \qty{6.0}{\tesla\metre} across most of the detector.
The muon spectrometer includes a system of precision tracking chambers up to \(|\eta| = 2.7\) and fast detectors for triggering up to \(|\eta| = 2.4\).
The luminosity is measured mainly by the LUCID--2~\cite{LUCID2} detector which is located close to the beampipe.
A two-level trigger system is used to select events~\cite{TRIG-2016-01}.
The first-level trigger is implemented in hardware and uses a subset of the detector information to accept events at a rate below \qty{100}{\kHz}.
This is followed by a software-based trigger that reduces the accepted event rate to \qty{1}{\kHz} on average depending on the data-taking conditions.
A software suite~\cite{SOFT-2022-02} is used in data simulation, in the reconstruction and analysis of real and simulated data, in detector operations, and in the trigger and data acquisition systems of the experiment.

\section{Simulation of signal and background processes}
\label{sec:simulation}

Samples of simulated events are used to model \ttjets production and most background processes.
These \ac{MC} simulated samples were generated employing either the full ATLAS detector simulation~\cite{SOFT-2010-01} based on \GEANT{}~\cite{Agostinelli:2002hh}, or a faster simulation where the \GEANT{} simulation of the calorimeter response is replaced by a detailed parameterization of the shower shapes~\cite{SOFT-2010-01}.
Both simulation methods were found to provide similar modeling for the observables used in this Letter.
To account for the effects of multiple interactions in the same and neighboring bunch crossings (pileup), additional interactions were simulated using \PYTHIA{}\,8.186~\cite{Sjostrand:2007gs} with a set of tuned parameters (\emph{tune}) referred to as A3~\cite{ATL-PHYS-PUB-2016-017} and superimposed onto the simulated hard-scatter event.
Subsequently, simulated events were reweighted to replicate the pileup conditions observed in the full \RunTwo data sample, with a mean number of \pp interactions per bunch crossing of \num{34}.
All simulated events were processed through the same reconstruction algorithms and analysis chain as the data.

Unless specified otherwise, \PYTHIA{8}~\cite{Sjostrand:2014zea} was employed to simulate \ac{PS}, hadronization, and \acp{MPI}.
For all samples using \PYTHIA{8} or \HERWIG{7}~\cite{Bahr:2008pv,Bellm:2015jjp,Bellm:2017jjp} for the simulation of the \ac{PS}, hadronization and \acp{MPI}, the decays of $b$- and $c$-hadrons were simulated using the \EVTGEN{} program~\cite{Lange:2001uf}.
\PYTHIA{8} setups use the A14 tune~\cite{ATL-PHYS-PUB-2014-021} and the \NNPDF[2.3lo] \ac{PDF} set~\cite{Ball:2012cx};
\HERWIG{7} setups use the H7UE tune~\cite{Bellm:2015jjp} or the default \HERWIG{} tune alongside the \MMHT[lo] \ac{PDF} set~\cite{Harland-Lang:2014zoa}.
The top-quark mass was set to $m_t = \qty{172.5}{\GeV}$ and, unless stated otherwise, the \ac{5FS} with massless \bquarks in the matrix element was used for all simulation setups and the corresponding \ac{PDF} sets.

Inclusive \ttbar events were generated with the \POWHEGBOX{2}~\cite{Frixione:2007nw,Nason:2004rx,Frixione:2007vw,Alioli:2010xd} generator at \ac{NLO} in \ac{QCD} employing the \NNPDF[3.0nlo] \ac{PDF} set~\cite{Ball:2014uwa}.
The \hdamp parameter, which regulates the transverse momentum (\pT) of the first additional emission beyond the Born configuration, was set to $1.5\,m_t$~\cite{ATL-PHYS-PUB-2016-020}.
The hardness scale parameter, which determines the region of phase space vetoed during showering when matched to a \ac{PS}, was fixed at $\pthard = 0$~\cite{Hoche:2021mkv,ATL-PHYS-PUB-2023-029}.
The scales \muR and \muF were set to the transverse mass of the top quark, $m_{\mathrm{T}}(t) = \sqrt{m_t^2 + p_{\mathrm{T},t}^2}$.
In the following, this sample is referred to as \ttbar \POWHEG{}+\PYTHIA{8}.
To assess uncertainties in the modeling, alternative sets of \ttbar events were generated with different configurations:
with $\pthard = 1$;
with $\hdamp = 3\,m_t$;
and with \POWHEGBOX{} interfaced to \HERWIG{7} instead of \PYTHIA{8}.
To compare the result with another prediction, an additional set was generated with the \MGNLO{}\,2.6.0 generator~\cite{Alwall:2014hca} at \ac{NLO} in \ac{QCD}, which uses the \NNPDF[3.0nlo] \ac{PDF} set, \MADSPIN{}~\cite{Frixione:2007zp,Artoisenet:2012st} to simulate top-quark decays, and \HERWIG{7} for \ac{PS} and hadronization~\cite{ATL-PHYS-PUB-2017-007}.
All generated sets of events were reweighted to the predicted \xsec of $\sigma(\ttbar) = 832 \pm 51 \;\unit{\pb}$, as calculated with the Top++2.0 program to \ac{NNLO} in QCD, including soft-gluon resummation to next-to-next-to-leading-log order (see Ref.~\cite{Czakon:2011xx} and references therein).
The uncertainties include independent variations of \muR and \muF, and variations in the \ac{PDF} and $\alpha_S$, following the \PDFforLHC prescription with the \MSTW 68\% CL \ac{NNLO}, \CT[10] \ac{NNLO} and \NNPDF[2.3] \ac{5FS} \ac{PDF} sets (see Ref.~\cite{Butterworth:2015oua} and references therein, and Refs.~\cite{Martin:2009bu,Gao:2013xoa,Ball:2012cx}).

The \POWHEGBOXRES{}~\cite{Jezo:2018yaf} generator and \OPENLOOPS{}~\cite{Cascioli:2011va,Denner:2016kdg,Buccioni:2019sur} were used to generate \ttbb events in the \ac{4FS} with massive \bquarks, where the generation of the additional \bbbar pair is included in the matrix element, employing the \NNPDF[3.0nlo] \ac{4FS} \ac{PDF} set~\cite{Ball:2014uwa}.
The \muR scale was set to $\frac{1}{2}\sqrt[4]{\prod_{i=t,\bar{t},b,\bar{b}}\,m_{\mathrm{T},i}}$ where $m_{\mathrm{T},i}$ denotes the transverse mass for each parton $i$.
The \muF scale and the \hdamp parameter were set to $0.5 \times \sum_{i} m_{\mathrm{T},i}$ with $i \in \{t,\bar{t},b,\bar{b},j\}$ and $i \in \{t,\bar{t},b,\bar{b}\}$, respectively, where $j$ denotes extra partons.
The \pthard parameter was fixed to 0, and the Born-zero-damp parameter (\hbzd), regulating the division between the finite and singular part of the real emission in the \ac{NLO} calculation, was set to 5.
In the following, this sample is referred to as \ttbb \POWHEG{}+\PYTHIA{8}.
To assess uncertainties in the modeling, alternative sets of \ttbb events were generated with different configurations:
with $\pthard = 1$;
using the dipole recoil scheme instead of the nominal global recoil scheme~\cite{Jezo:2018yaf,ATL-PHYS-PUB-2022-026};
and interfacing \POWHEGBOXRES{} with \HERWIG{7} instead of \PYTHIA{8}.
To compare the result with another prediction, alternative sets of \ttbb events were generated with $\hbzd = 2$.
An additional alternative set of events was generated with the \SHERPA{}\,2.2.10 generator~\cite{Bothmann:2019yzt} in the \ac{4FS}, for which the virtual corrections for matrix elements at \ac{NLO} accuracy were provided by \textsc{Comix}~\cite{Gleisberg:2008fv} and \OPENLOOPS.
The \SHERPA{} events were matched with the \SHERPA{} \ac{PS} algorithm~\cite{Schumann:2007mg} based on Catani--Seymour dipole factorization using the MEPS@NLO prescription~\cite{Hoeche:2011fd,Hoeche:2012yf,Catani:2001cc,Hoeche:2009rj} and a set of tuned parameters developed by the \SHERPA{} authors.
All generated sets of events were reweighted to the same prediction \xsec value as computed in \POWHEGBOXRES at \ac{NLO} in \ac{QCD}.

A dedicated simulation of \ttcc production in the \ac{3FS} is not available.
Hence, \ttcc contributions are estimated solely using \ttbar \ac{5FS} simulation, where only the gluon radiation process is simulated in the \ac{NLO} matrix element, but its splitting into a \ccbar pair is done in the \ac{PS}.

Simulated \ttbar and \ttbb events are sorted into four \ttjets categories using particle-level information: \ttccin, \ttc, \ttbin, and \ttlight.
For the categorization, jets are reconstructed from stable particles\footnote{Particle-level objects are considered stable if $\tau > \qty{3e-11}{\s}$.}
following \ac{PS} and hadronization, using the \antikt algorithm~\cite{Cacciari:2008gp,Fastjet} with a radius parameter $R = 0.4$.
They are required to have $\pt > \qty{15}{\GeV}$ and $\abseta < 2.5$.
The flavor of a particle-level jet is determined by counting $b$- and $c$-hadrons with $\pt > \qty{5}{\GeV}$ that are \emph{ghost-associated}~\cite{Cacciari:2007fd,Cacciari:2008gn} with the jet.
Jets with one or more associated $b$-hadrons are designated as \bjets, while those with one or more associated $c$-hadrons but no $b$-hadron are labeled as \cjets.
Events are categorized based on the presence of additional particle-level \bjets and \cjets,
excluding jets originating from decays of top quarks and \Wboson~bosons.\footnote{%
The procedure to exclude these jets is as follows:
\bquarks and \cquarks originating directly from the decays of top quarks and \Wboson~bosons are identified in the \ac{MC} generator record.
Then, the $b$- and $c$-hadrons closest in $\Delta R$ are flagged.
Any jet \emph{ghost-associated} with these hadrons is removed from the categorization.
}
Events with one or more \bjets are classified as \ttbin, those with two or more \cjets but no \bjet are categorized as \ttccin, and events with one \cjet but no \bjet are labeled as \ttc.
Any remaining events enter the \ttlight category.

The \ttjets categorization at particle level is used to remove the overlap between the inclusive \ttbar \ac{5FS} simulation and the \ttbb \ac{4FS} simulation.
The latter is expected to provide the more accurate modeling of \ttbb production and is the preferable setup to simulate \ttbin contributions in this analysis.
Thus, events are removed from any of the \ttbar \ac{5FS} setups if they fall into the \ttbin category at particle level.
All other events are retained to estimate contributions in the \ttccin, \ttc, and \ttlight categories.
Likewise, events from the \ttbb \ac{4FS} simulation are removed if they fall into the \ttccin, \ttc, or \ttlight categories at particle level.
The individual \xsecs assigned to the \ttbar and \ttbb samples are retained, and no rescaling is performed after removing the overlapping events.

Various other processes involving top quarks can mimic \ttjets topologies.
Single-top-quark $t$-channel, $s$-channel, and $tW$ production were simulated using \POWHEGBOX{2} at \ac{NLO} in \ac{QCD}.
For $t$-channel production, events were generated in the \ac{4FS} with the \NNPDF[3.0nlo] \ac{4FS} \ac{PDF} set.
For $s$-channel and $tW$ production, events were generated in the \ac{5FS} using the \NNPDF[3.0nlo] \ac{5FS} \ac{PDF} set.
The $tW$ simulation employed the diagram-removal scheme~\cite{Frixione:2008yi} to treat interference with \ttbar production~\cite{ATL-PHYS-PUB-2016-020}.
Additional sets of $tW$ events were generated in different setups:
using the diagram-subtraction scheme~\cite{Frixione:2008yi,ATL-PHYS-PUB-2016-020};
using \POWHEGBOX{2} interfaced with \HERWIG{}\,7.04;
and using \MGNLO{}\,2.6.2 interfaced with \PYTHIA{8}.
\POWHEGBOX{2}+\HERWIG{7} setups were also generated for $t$-channel and $s$-channel production.
\ttH production was simulated with \POWHEGBOX{2} at \ac{NLO} in \ac{QCD} in the \ac{5FS}, using the \NNPDF[3.0nlo] \ac{PDF} set.
Additionally, \ttW, \ttZ, $tZq$, and $tWZ$ events were generated using the \MGNLO{}\,2.3.3 generator at \ac{NLO} in \ac{QCD} with the \NNPDF[3.0nlo] \ac{PDF} set.
Secondary sets of \ttW and \ttZ events were simulated with \SHERPA{}\,2.2.0 at \ac{LO} in \ac{QCD}.
In the following, these processes are collectively referred to as \emph{Other Top}.

\Wjets and \Zjets events were simulated using the \SHERPA{}\,2.2.1 generator employing \ac{NLO} matrix elements for up to two partons and \ac{LO} matrix elements for up to four partons, calculated with the \textsc{Comix} and \OPENLOOPS{} libraries.
Diboson events with semileptonic and fully leptonic decays were simulated using \SHERPA{}\,2.2.1 and \SHERPA{}\,2.2.2, employing \ac{NLO} matrix elements for up to one additional parton and \ac{LO} matrix elements for up to three additional partons.
\Wjets, \Zjets and diboson events were matched with the \SHERPA{} \ac{PS} algorithm based on Catani--Seymour dipole factorization using the MEPS@NLO prescription and a set of tuned parameters developed by the \SHERPA{} authors.
The \NNPDF[3.0nnlo] \ac{PDF} set~\cite{Ball:2014uwa} was used, and the \xsecs of the samples were reweighted to \ac{NNLO} predictions~\cite{Anastasiou:2003ds}.
In the following, \Wjets, \Zjets and diboson processes are collectively referred to as \emph{Non-Top}.

The \xsec measurements are conducted within a fiducial volume at particle level, closely resembling the detector-level acceptance, ensuring robustness against variations in acceptance predictions from \ac{MC} simulations.
Events in the \slepton (\dilepton) channel must exhibit five (three) or more jets with $\pt > \SI{25}{\GeV}$ and $\abseta < 2.5$, out of which at least two must be ghost-associated with $b$-hadrons.
No specific requirements are imposed regarding the presence of additional $b$-~or \cjets in the fiducial phase space, but the \ttjets categorization is performed as described above.
In the \slepton channel, events must feature exactly one charged lepton ($\ell = e, \mu$), while in the \dilepton channel, events must contain exactly two oppositely charged leptons.
One lepton must have $\pT > \qty{27}{\GeV}$, any additional lepton must carry $\pT > \qty{10}{\GeV}$, and all are required to be within $\abseta < 2.5$.
Any event featuring a same-flavor lepton pair with an invariant mass below \qty{15}{\GeV} or within the range \SIrange{83}{99}{\GeV} is rejected.
Leptons are removed if their distance to a jet satisfies $\Delta R < 0.4$.
The fiducial phase space accepts approximately \SI{17}{\percent} of the total number of simulated \ttcin events.
The measurements are also performed in a more inclusive phase-space volume without requirements on the \ttbar decay products and the jet multiplicity.

\section{Event reconstruction and selection}

Events are required to have at least one primary vertex with two or more tracks with $\pt > \SI{0.5}{\GeV}$.
For events with more than one primary vertex, the hard-scattering primary vertex is selected as the one with the highest sum of squared track \pt~\cite{ATL-PHYS-PUB-2015-026}.
A suite of single-electron and single-muon triggers is used to select events with at least one charged lepton~\cite{TRIG-2018-05,TRIG-2018-01}.

Jets are reconstructed using the \antikt clustering algorithm with a radius parameter of $R=0.4$ using particle flow jet constituents that combine measurements from both the \ac{ID} and the calorimeter~\cite{PERF-2015-09}.
Jet candidates undergo calibration using simulations, with corrections derived from in situ techniques applied to data~\cite{JETM-2018-05}.
They are required to have $\pt > \SI{25}{\GeV}$ and $\abseta < 2.5$, with jets within $\abseta < 2.4$ and $\pt < \SI{60}{\GeV}$ further required to pass the tight working point of the jet vertex tagger~\cite{PERF-2014-03} to reduce contribution from pileup jets.

Electron candidates are reconstructed from clusters of energy in the \ac{EM} calorimeter associated with reconstructed tracks from the \ac{ID}.
Their reconstruction, identification, and calibration are detailed in Refs.~\cite{EGAM-2018-01,EGAM-2021-01}.
They must satisfy a set of likelihood-based identification criteria with $\pt > \SI{10}{\GeV}$ and $\absetacluster < 2.47$.
Electrons in the transition region between the end-caps and barrel region ($1.37 < \absetacluster < 1.52$) are vetoed.
Tracks matched to electrons are required to be associated with the primary vertex and satisfy $|z_0\sin\theta| < \SI{0.5}{\mm}$ and $|d_0|/\sigma(d_0) < 5$, where $z_0$ and $d_0$ are the longitudinal and transverse impact parameters of the electron track, respectively.
For the analysis regions, the identification criteria are tightened, and electrons are required to satisfy a set of variable-radius isolation criteria to reduce contributions from misidentified jets, considering energy depositions in the calorimeter and tracks in the \ac{ID}~\cite{EGAM-2018-01}.

Muons are reconstructed by combining a track from the muon spectrometer with an \ac{ID} track.
Details regarding the reconstruction, identification and calibration of muons are summarized in Refs.~\cite{MUON-2018-03,MUON-2022-01}.
The \ac{ID} tracks must be associated with the primary vertex by passing $|z_0\sin\theta| < \SI{0.5}{\mm}$ and $|d_0|/\sigma(d_0) < 3$.
They must fulfill $\pt > \SI{10}{\GeV}$, $\abseta < 2.5$, and satisfy a set of muon quality criteria.
For the analysis regions, the quality criteria are tightened, and muons are required to pass a track-based isolation criterion to reduce contributions from misidentified jets~\cite{MUON-2018-03}.

An overlap removal procedure prevents double counting of energy deposits and improves object resolution and identification efficiencies.
Electron candidates sharing tracks with a muon candidate are removed.
The nearest jet within $\Delta R < 0.2$ of an electron candidate is removed in favor of the electron.
Electrons within $\Delta R < 0.4$ of any jets after this selection are removed in favor of the jet.
Muons are removed if they are within $\Delta R < 0.4$ of a jet to reduce contributions from heavy-flavor decays.
However, if the jet has fewer than three associated tracks, the muon is kept in favor of the jet.

Jets containing $b$-hadrons are identified as \btagged jets using the DL1r algorithm~\cite{FTAG-2019-07}, which leverages distinctive $b$-hadron features like track impact parameters and the presence of displaced vertices in the \ac{ID}.
The algorithm also includes discriminating variables from a recurrent neural network that exploit spatial and kinematic correlations between tracks from the same $b$-hadron.
The DL1r algorithm provides three output scores, $p_b$, $p_c$, and $p_{\text{light}}$, and in the standard DL1r calibration, jets are tagged as \bjets using a discriminant $\mathcal{D}_b$ that is built using the Neyman--Pearson lemma~\cite{doi:10.1098/rsta.1933.0009}:
\begin{equation*}
\mathcal{D}_b = \mathrm{log}\frac{p_b}{f_c p_c+(1-f_c) p_{\text{light}}}.
\end{equation*}
The parameter~$f_c$ controls which background contributes more to the decision.
To improve the efficiency of selecting jets originating from \cquarks without compromising the \bjet identification performance, the DL1r algorithm was reoptimized as a 2D binned discriminant, the \bctagger.
This reoptimization was necessary because standard DL1r configurations do not provide calibrated \ctag working points, requiring a tailored approach to achieve the desired balance between \bjet and \cjet identification.
The axes of the discriminant, $\mathcal{D}_c$ and $\mathcal{D}_b$, are calculated from the $p_b$, $p_c$, and $p_{\text{light}}$ scores following the above formula, with all $b$- and $c$-subscripts interchanged for the $\mathcal{D}_c$ discriminant.
It was found that $f_b = 0.4$ provides good performance for the \bctagger as it leads to a balanced contamination of \bjets and light-flavor jets in the analysis regions; the value of $f_c = 0.018$ was chosen to align the admixture with the standard DL1r calibration.

\begin{figure}[tp]
\centering
\includegraphics[width=0.6\textwidth]{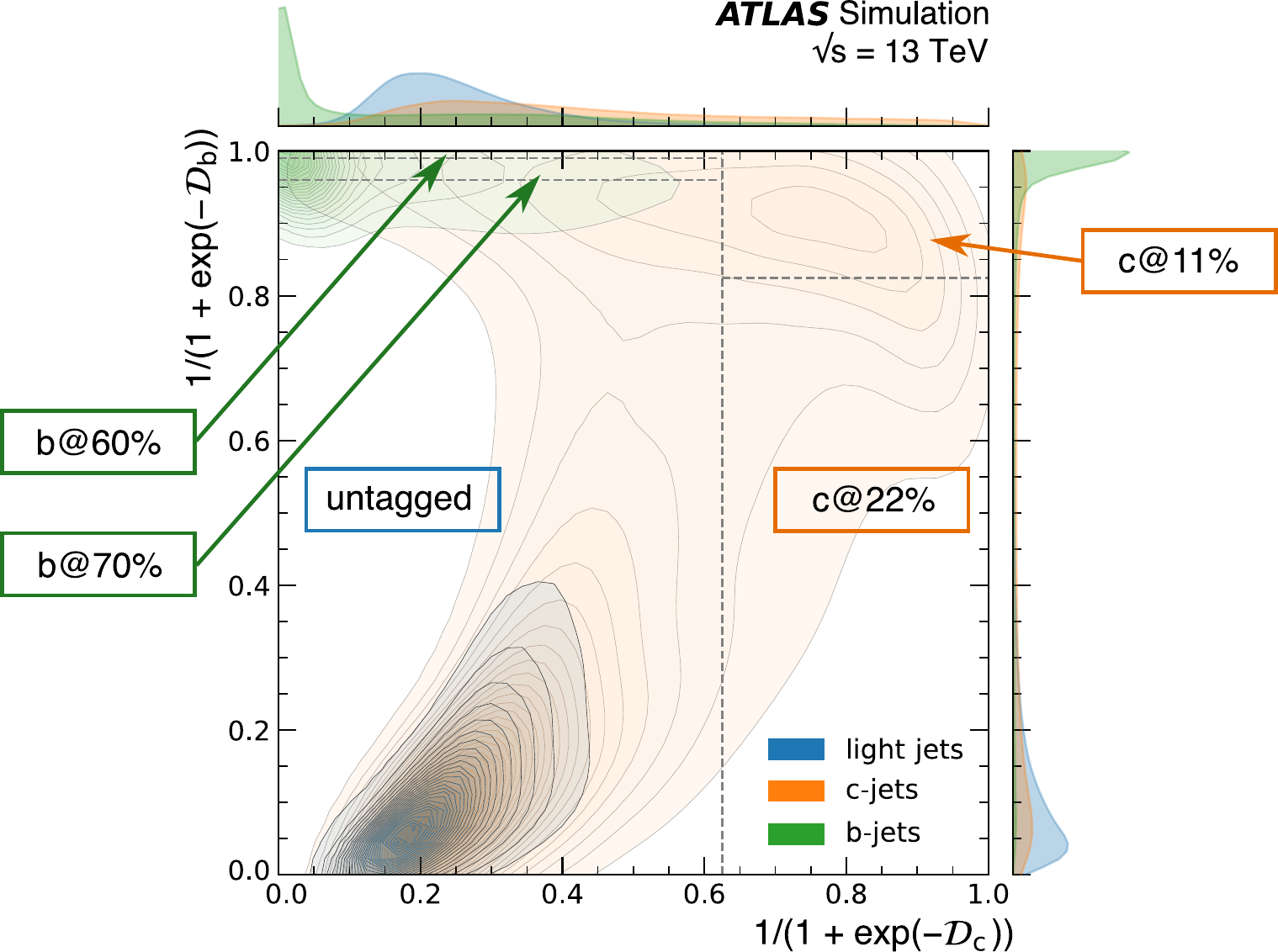}
\caption{Distribution of light, $c$-, and $b$-jets for the 2D \bctagger in simulated \ttbar events.
Dashed lines correspond to the edges of the working points.
For visualization purposes a standard logistic function is applied to both axes of the discriminant.
The contours for each jet type are smoothed with a kernel density method to improve readability, with contours corresponding to lines of constant density.
The two \btag bins comprise the top left corner of the discriminant, with the \ctag bins to the right of the vertical dashed line.
Untagged jets are located in the bottom left corner.
}
\label{fig:bctagger}
\end{figure}

The binning in $\mathcal{D}_c$ and $\mathcal{D}_b$ includes five \acp{WP}, two of which are optimized for \cjet identification and two for \bjet identification.
The \acp{WP} are designed to achieve an approximately constant efficiency for \bjet and \cjet identification across the jet \pT and $\abseta$ range, with $\pT > \SI{20}{\GeV}$.
The \ctag \acp{WP} are separated from the others by $\mathcal{D}_c' \geq 0.625$, where the prime indicates that a standard logistic function is applied to the discriminant: $\mathcal{D}_c' = 1/(1+\exp(-\mathcal{D}_c))$.
The tight \ctag \ac{WP} includes a tighter cut on $\mathcal{D}_b'$ to increase the light-flavor rejection rate and is designed to achieve an exclusive efficiency of 11\% (denoted \bctightc).\footnote{%
Tightening the cut on $\mathcal{D}_c'$ instead of $\mathcal{D}_b'$ would lead to a more balanced increase in the \bjet and light-flavor rejection rate, but it was found that a focus on higher light-flavor rejection rates yields better \ttcin signal sensitivity.
}
The loose \ctag \ac{WP}, which includes \bctightc, achieves an inclusive efficiency of 22\% (\bcloosec).
The \bctightc and \bcloosec \acp{WP} yield \bjet rejection rates of 28.7 and 18.9 and light-flavor rejection rates of 1051 and 104 in simulated \ttbar events, respectively.
The \btag \acp{WP}, defined for $\mathcal{D}_c' < 0.625$, use the same $\mathcal{D}_b'$ cuts as the two tightest \btag \acp{WP} in the standard DL1r calibration.
The tighter \btag \ac{WP} is designed to achieve an exclusive efficiency of 60\% (\bctightb),
the looser, which includes \bctightb, to achieve an inclusive \btag efficiency of 70\% (\bclooseb),
with \cjet rejection rates of 37.1 and 12.2 and light-flavor rejection rates of 2320 and 573, respectively.
Jets not entering any of the four \ac{WP} bins are classified as untagged.
Consequently, the \bctagger comprises a total of five orthogonal bins, with the distribution of light, $c$- and \bjets illustrated in \cref{fig:bctagger}.

The calibration of the five \bctagger \acp{WP} follows standard procedures applied to the DL1r algorithm~\cite{FTAG-2018-01,FTAG-2020-08,FTAG-2019-02}.
Jets originating from $b$- and \cquarks are calibrated in data using \ttbar events, both selections being orthogonal to the events analyzed in this study by limiting the number of jets to four (two) in \slepton (\dilepton) final states.
All \btag calibration \acp{SF} are close to unity, with statistical uncertainties dominating the high \pT regime.
The \ctag calibration yields more variation in the \ac{SF} values.
For the \bclooseb \ac{WP}, the \acp{SF} average around 0.9 and some bins in jet \pT are incompatible with unity within their uncertainties.
Light-flavor jet calibration employs the \emph{negative-tag} method~\cite{ATLAS-CONF-2014-046,D0:2006hcm} in conjunction with a \emph{flip-tagger} approach following Ref.~\cite{FTAG-2019-02}, and all \acp{SF} are compatible with unity within uncertainties.
For the \bctightb working point, the light-flavor \acp{SF} cannot be determined due to the large \bjet contamination and so, following the procedure in Ref.~\cite{FTAG-2019-02}, are set to unity with additional uncertainties.
The \bctightc \ac{WP} also has large uncertainties due to high contamination of heavy-flavor jets and low statistics.
Simulation-to-simulation \acp{SF} are derived in bins of jet \pT and \abseta to account for differences between tagging efficiencies for different \ac{PS} and hadronization algorithms~\cite{ATL-PHYS-PUB-2020-009}.

A preselection is applied and requires at least one electron or muon with $\pt > \SI{27}{\GeV}$ that is matched to the trigger object.
The \slepton channel selects events with exactly one lepton, while the \dilepton channel selects those with exactly two oppositely charged leptons.
Events must have at least five (three) jets in the \slepton (\dilepton) channel, with at least three (two) \btagged or \ctagged jets using the \bclooseb or \bcloosec working points.
For dilepton final states with identical lepton flavors, the same invariant-mass criteria as those applied in the fiducial selection at particle level are imposed.
In the \slepton channel, events are split into those with exactly five jets (5-jet-exclusive) and those with six or more jets (6-jet-inclusive).
Similarly, in the \dilepton channel, events are categorized as either 3-jet-exclusive or 4-jet-inclusive.
Based on the number of identified $b$- and \ctagged jets, events are further classified into \acp{SR} enriched in \ttccin and \ttc events, and \acp{CR}, which determine the normalization of the \ttbin and \ttlight background.
An overview of the seven \acp{SR} and 12 \acp{CR} is shown in \cref{tab:regions}.
Approximately \SI{5.3}{\percent} of the simulated \ttcin events passing the fiducial selection criteria are reconstructed in the analysis regions.

\begin{table}[tp]
\centering
\caption{
Jet multiplicity and tagging criteria for the \acfp{SR} and \acfp{CR} in the \slepton and \dilepton channels.
All \slepton regions are defined in the 5-jet-exclusive and 6-jet-inclusive jet selections, all \dilepton regions in the 3-jet-exclusive and 4-jet-inclusive jet selections.
By construction, \srtightll only exists for the 4-jet-inclusive selection.
Identical requirements on inclusive and exclusive working points, e.g., one \bcloosec and one \bctightc in the \crOnel, indicate a veto of additional tags at the inclusive working point.
}
\label{tab:regions}
\small
\begin{tabular}{l|ccccc|cccc|c}
\toprule
& \crOnel & \crTwol & \crThreel & \srloosel & \srtightl & \crOnell  & \crTwoll & \crThreell  & \srloosell & \srtightll \\
\midrule
$N_\mathrm{jets}$ & \multicolumn{5}{c|}{$=5$ or $\geq 6$} & \multicolumn{4}{c|}{$=3$ or $\geq 4$} & $\geq$ 4 \\
\midrule
\bclooseb & 2  & --        &  -- & 2         & 2         & 2   & --         & \(\geq\) 3 & 2  & 2 \\
\bctightb & -- & \(\geq3\) &  3  & --        & --        & --  & \(\geq\) 3 & \(\leq\) 2 & -- & -- \\
\bcloosec & 1  & 0         &  1  & \(\geq2\) & --        & 0   & --         & --         & 1  & \(\geq2\) \\
\bctightc & 1  & --        &  1  & 1         & \(\geq2\) & --  & --         & --         & -- & -- \\
\bottomrule
\end{tabular}
\end{table}

There are four \acp{SR} in the \slepton channel and three \acp{SR} in the \dilepton channel with different degrees of purity in \ttccin and \ttc events.
The \srtightl and \srtightll regions target \ttccin events, while \srloosel and \srloosell are dominated by \ttc.
All seven regions require exactly two \btagged jets at \bclooseb.
In the \slepton channel, events featuring at least two \ctagged jets at \bcloosec with exactly one satisfying the \bctightc working point enter the \srloosel regions.
Events with two or more \ctagged jets at \bctightc are classified into \srtightl.
The expected purity in \ttccin events in the \slepton \acp{SR} ranges between 10\% and 28\%, while the predicted \ttc contributions are between 29\% and 38\%.
In the \dilepton channel, events with exactly one \ctagged jet at \bcloosec are sorted into \srloosell, while events with at least two enter \srtightll, which is only defined in the 4-jet-inclusive selection.
In the \dilepton regions, the expected purity in \ttccin and \ttc events is between 4\% and 46\%, and between 16\% and 38\%, respectively.

In the \slepton channel, three \acp{CR} with varying compositions in \ttbin and \ttlight are defined for each of the two jet multiplicity selections.
The \crOnel regions select exactly two \btagged jets at \bclooseb and exactly one \ctagged jet passing \bcloosec and \bctightc.
They are dominated by \ttlight events.
The \crTwol regions are characterized by events with three or more \btagged jets at \bctightb, with a veto on \ctagged jets, leading to a mix of \ttbin and \ttlight.
The \crThreel regions are pure in \ttbin and select events with exactly three \btagged jets at \bctightb and exactly one \ctagged jet satisfying \bcloosec and \bctightc.

Similarly, the \dilepton channel features three \acp{CR} for each of the two jet multiplicity selections.
The \crOnell regions require exactly two \btagged jets at \bclooseb and veto \ctagged jets.
They are dominated by \ttlight events.
The \crTwoll regions are pure in \ttbin and select events with three or more \btagged jets at \bctightb.
The \crThreell regions show a mix of \ttbin and \ttlight by requiring at least three \btagged jets at \bclooseb, with no more than two satisfying the \bctightb criteria.

Contributions from processes with non-prompt or misidentified leptons to the \slepton selection, in the following termed \emph{fake leptons}, were estimated by using the data-driven \emph{matrix method}~\cite{EGAM-2019-01} in dedicated \acp{CR}.
Prompt-lepton efficiencies were derived in data using a tag-and-probe method in \Zboson-boson decays.
Fake-lepton efficiencies were estimated in data in bins of lepton \pT and \abseta using events with at least three jets, at least two \btagged jets at the 70\% working point of the DL1r tagger, and one lepton fulfilling the looser identification and quality criteria.
The isolation requirements were completely dropped.
To enrich the selection in fake leptons, only events where the scalar sum of the magnitude of the missing-transverse-momentum vector and the leptonically decaying \Wboson-boson mass is no larger than \SI{60}{\GeV} were considered.\footnote{%
The magnitude of the missing-transverse-momentum vector, denoted \MET, is defined as the negative sum of the transverse momenta of the reconstructed and calibrated physical objects, plus a \emph{soft term} built from all other tracks associated with the primary vertex~\cite{JETM-2020-03} and not matched to a reconstructed object.
The mass of the leptonically decaying \Wboson~boson is calculated from the kinematics of the lepton and the missing transverse momentum.
}
The fake-lepton contributions to the analysis were then estimated by inverting the efficiency matrix and applying it to the observed event yields in the analysis regions using the looser and the nominal lepton identification and isolation criteria.
Fake-lepton contributions to the \dilepton selection mainly arise from \Wjets and \slepton \ttjets events and were estimated from \ac{MC} simulation.

\section{Systematic uncertainties}

The extraction of the \ttccin and \ttc cross-sections is subject to various sources of uncertainties, impacting either the overall event yields in a region, the shape of observable distributions considered in the fit, or both.

Experimental sources of uncertainties include uncertainties in the value of the integrated luminosity of the \RunTwo data, the simulation of pileup events, and effects related to the reconstruction and identification of physics objects used in the analysis.
The uncertainty in the combined 2015--2018 integrated luminosity is 0.83\% \cite{DAPR-2021-01}, obtained using the LUCID-2 detector~\cite{LUCID2} for the primary luminosity measurements, complemented by measurements using the inner detector and calorimeters.
Uncertainties in the modeling of pileup are evaluated by varying the pileup reweighting in \ac{MC} simulations within its associated uncertainties.
Furthermore, lepton identification and isolation efficiencies, momentum scale and resolution, and lepton trigger efficiencies are varied within their uncertainties to evaluate their impact on the measurement~\cite{MUON-2018-03,MUON-2022-01,EGAM-2018-01,EGAM-2021-01}.
The uncertainty in the \ac{JES} is derived from a combination of simulations, test-beam data, and in situ measurements~\cite{JETM-2018-05}.
This uncertainty incorporates contributions from various sources such as jet-flavor composition, $\eta$-intercalibration, punch-through, single-particle response, calorimeter response to different jet flavors, and pileup.
It comprises 30 uncorrelated JES uncertainty subcomponents.
Additionally, the jet energy resolution in simulation is varied by its corresponding uncertainty, which is divided into thirteen uncorrelated sources.
The uncertainty associated with the jet vertex tagger is determined by varying its efficiency correction factors~\cite{PERF-2014-03}.
Moreover, uncertainties in the calibration of the \bctagger are separately determined for \bjets, \cjets, and light-flavor jets, following the procedures detailed in Refs.~\cite{FTAG-2018-01,FTAG-2020-08,FTAG-2019-02}.
They are derived as a function of jet \pT and separately for each \ac{WP}, yielding a total of 45 components for \bjets and 20 each for \cjets and light-flavor jets.
For jets with \pT values above the range covered by the calibration, extrapolation uncertainties derived from \ac{MC} simulation are applied.

Uncertainties in the modeling of \ttccin, \ttc, and \ttlight are considered through several alternative inclusive \ttbar simulation setups.
The alternative set of events with $\pthard = 1$ is used to assess the uncertainty in the \ac{NLO} matching between the matrix elements and the \ac{PS}.
The $\hdamp = 3\,m_t$ sample assesses the uncertainty in the choice of that parameter.
The \POWHEG{}+\HERWIG{7} setup evaluates the uncertainty in the choice of the \ac{PS} and hadronization algorithm.
Additionally, the nominal set of events is reweighted to scenarios where \muR and \muF in the matrix element are halved and doubled independently to assess the uncertainty in the choice of these parameters.
To estimate uncertainties in the modeling of \ac{ISR} and \ac{FSR}, the factorized parameter $\alpha_S^{\mathrm{ISR}}$ is varied using the \emph{var3c} variation of the A14 tune, and the renormalization scale associated with $\alpha_S^{\mathrm{FSR}}$ is varied to 0.625 and 2 relative to its nominal setting.
The uncertainty in the \ac{PDF} set is estimated by using 30 \ac{PDF} variations of the \PDFforLHC prescription.

Uncertainties in the modeling of the \ttbin contributions are considered through a similar suite of alternative \ttbb simulation setups.
The alternative set of events with $\pthard = 1$ is used to assess the uncertainty in the \ac{NLO} matching between the matrix elements and the \ac{PS}.
The uncertainty in the choice of the recoil scheme is evaluated with the set of events generated with the alternative dipole recoil scheme.
The \POWHEG{}+\HERWIG{7} setup is used to evaluate the uncertainty in the selection of the \ac{PS} and hadronization algorithm.
Event weight variations identical to those applied in \ttbar simulations, including \muR, \muF, \ac{ISR}, \ac{FSR}, and the \ac{PDF} set, are considered for \ttbb.

All uncertainties considered in the modeling of \ttjets production are treated as uncorrelated between the \ttbin, \ttcin, and \ttlight processes.
Moreover, the \ac{NLO} matching and \ac{PS} uncertainties for \ttcin (\ttbin) are parameterized independently for the \ttccin and \ttc (\ttbbin and \ttb) components as the components are sensitive to different effects.
Additionally, the \ttlight \ac{PS} uncertainty is parameterized separately for the 3-jet-exclusive, 4-jet-inclusive, 5-jet-exclusive, and 6-jet-inclusive regions to avoid strong constraints arising from region-to-region migration effects.

For $t$-channel, $s$-channel, and $tW$ single-top-quark production, the \POWHEG{}+\HERWIG{7} setups are used to assess the uncertainty in the choice of the \ac{PS} and hadronization algorithm.
For $tW$ production, the set of events with the diagram-subtraction scheme is used to gauge the uncertainty in the treatment of the interference with \ttbar,
and the \ac{NLO} matching uncertainty is evaluated using events generated with \MGNLO{}+\PYTHIA{8}.
An additional 5\% normalization uncertainty is assigned to accommodate the uncertainty in the \xsec value~\cite{Kidonakis:2021vob}.
For the $t$-channel and $s$-channel predictions, normalization uncertainties of 50\% are assigned to cover NLO matching and \xsec uncertainties.
The difference between the sets of \ttW and \ttZ events generated with \MGNLO{}+\PYTHIA{8} and \SHERPA{} is used as an uncertainty in the prediction for these processes.
Additionally, uncertainties of approximately 13\% and 11\% are assigned to \ttW and \ttZ, respectively, to accommodate uncertainties in the assigned \xsec values~\cite{deFlorian:2016spz}.
Conservative normalization uncertainties of 50\% are assigned to each of the remaining top-quark processes considered (\ttH, $tWZ$, $tZq$) to cover potential mismodeling of the rates of these processes in the probed fiducial phase space.
For the \Wjets, \Zjets and diboson background, the assigned uncertainties follow those used in previous ATLAS measurements~\cite{HIGG-2020-03-corr}.
Normalization uncertainties of 40\% and 35\% are considered for \Wjets and \Zjets processes, based on variations of the \muR and \muF scales and of the matching parameters in the \SHERPA{} generator.
Additional 40\% uncertainties are assigned to \Wjets events with exactly two heavy-flavor jets and at least three heavy-flavor jets based on the same variations.
A 50\% uncertainty is assigned to the diboson background, which includes uncertainties in the inclusive \xsec and additional jet production~\cite{Grazzini:2016swo,ATL-PHYS-PUB-2016-002,STDM-2018-03}.
None of the modeling uncertainties of the \emph{Other Top} and \emph{Non-Top} categories have any significant impact on the \ttccin and \ttc cross-section measurements.

The fake-lepton estimate in the \slepton channel is derived from a data sample with finite statistics, leading to bin-by-bin statistical uncertainties for the fake-electron and fake-muon contributions.
In addition, conservative 50\% normalization uncertainties are assigned to the data-driven \slepton and \ac{MC}-based \dilepton estimates.

\section{Results}

The data from all 19 analysis regions are simultaneously fit using a binned profile likelihood approach using the HistFactory~\cite{histfactory} and RooFit~\cite{RooFit} toolkits.
Systematic uncertainties are incorporated as nuisance parameters in the fit and are constrained by a Gaussian penalty term present in the likelihood function.
The statistical uncertainty arising from the limited number of simulated events is included in the likelihood in the form of additional nuisance parameters with Poisson constraint terms.
To determine the production cross-sections of the \ttccin and \ttc processes, and to control the \ttbin and \ttlight background processes, normalization factors are defined for all four \ttjets categories.
These factors, $\mu_i$, serve as unconstrained normalization factors in the profile likelihood fit, and they are applied to both fiducial and non-fiducial \ttjets events.
The inclusive \ttjets normalization factor is computed following
\begin{align}
\muttjets = \sum_{i} \mu_i R_i^{\mathrm{MC}}\quad \text{for}~ i \in \{\ttbin, \ttccin, \ttc, \ttlight\},
\end{align}
where $R_i^{\mathrm{MC}}$ is the predicted \xsec ratio of process~$i$ to overall \ttjets production in \ac{MC} simulation.

Each of the 12 \acp{CR} is represented as a single bin in the fit.
\srlooselfivej and \srtightlfivej use the invariant mass between the two geometrically closest \ctagged jets to increase sensitivity to differences between the \ttccin, \ttc, and \ttlight contributions (only considering \ctagged jets passing \bctightc in \srtightlfivej).
Due to \cjets originating from the decay of the \Wboson~boson, the contribution from \ttlight is enhanced around the \Wboson-boson mass.
\srloosellthreej uses the invariant mass between the \ctagged jet and the geometrically closest \btagged jet.
All three regions use four bins with varying width to maximize sensitivity to the \ttccin and \ttc contributions.
In the 4-jet-inclusive and 6-jet-inclusive \acp{SR}, the jet multiplicity is used to enhance the separation between \ttbin, \ttccin, \ttc, and \ttlight processes, fitting $6 \leq N_\mathrm{jets} \leq 9$ and $4 \leq N_\mathrm{jets} \leq 7$ in the \slepton and \dilepton channels, respectively.
Overflow events are included in the last bin of each distribution.

\begin{figure}[tp]
\centering
\includegraphics[width=0.99\textwidth]{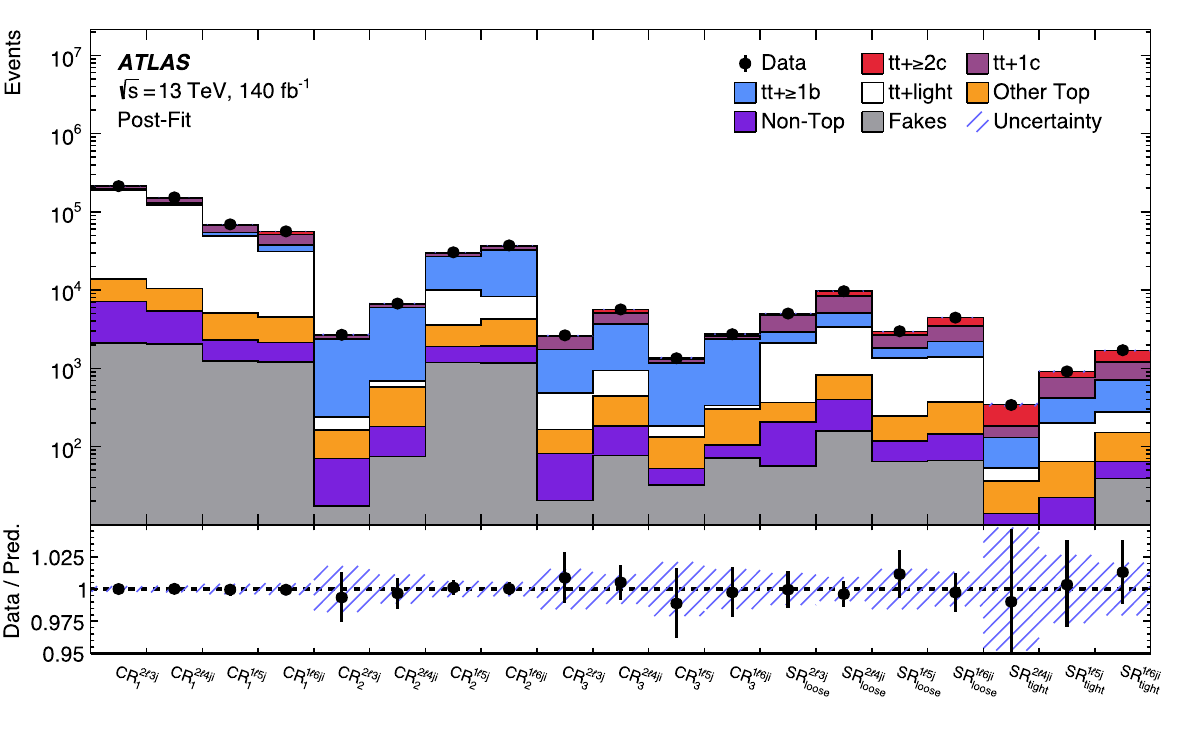}
\caption{%
Post-fit agreement between data and \ac{MC} simulation in the 12 \acfp{CR} and seven \acfp{SR}.
The error bars indicate data statistical uncertainties.
The hatched uncertainty bands include all uncertainties and their correlations.
\emph{Other Top} includes single-top-quark production and associated production of \ttbar and single top quarks with bosons.
\emph{Non-Top} includes \Wjets, \Zjets, and diboson processes.
}
\label{fig:post_fit_regions}
\end{figure}

\begin{figure}[p]
\centering
\subcaptionbox{\label{fig:post-fit-plot-5j-loose}}{\includegraphics[width=0.48\textwidth]{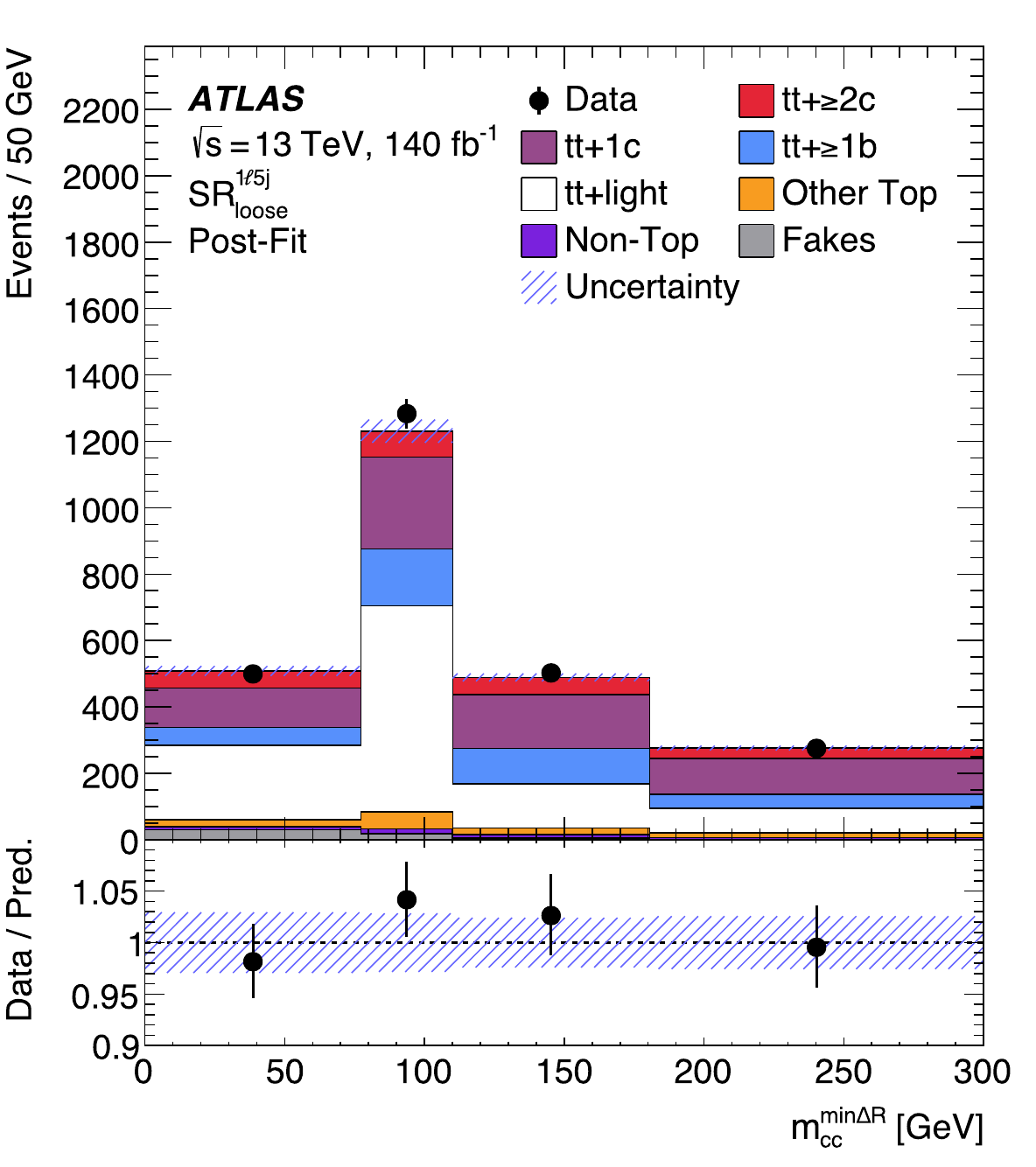}}\hfill
\subcaptionbox{\label{fig:post-fit-plot-5j-tight}}{\includegraphics[width=0.48\textwidth]{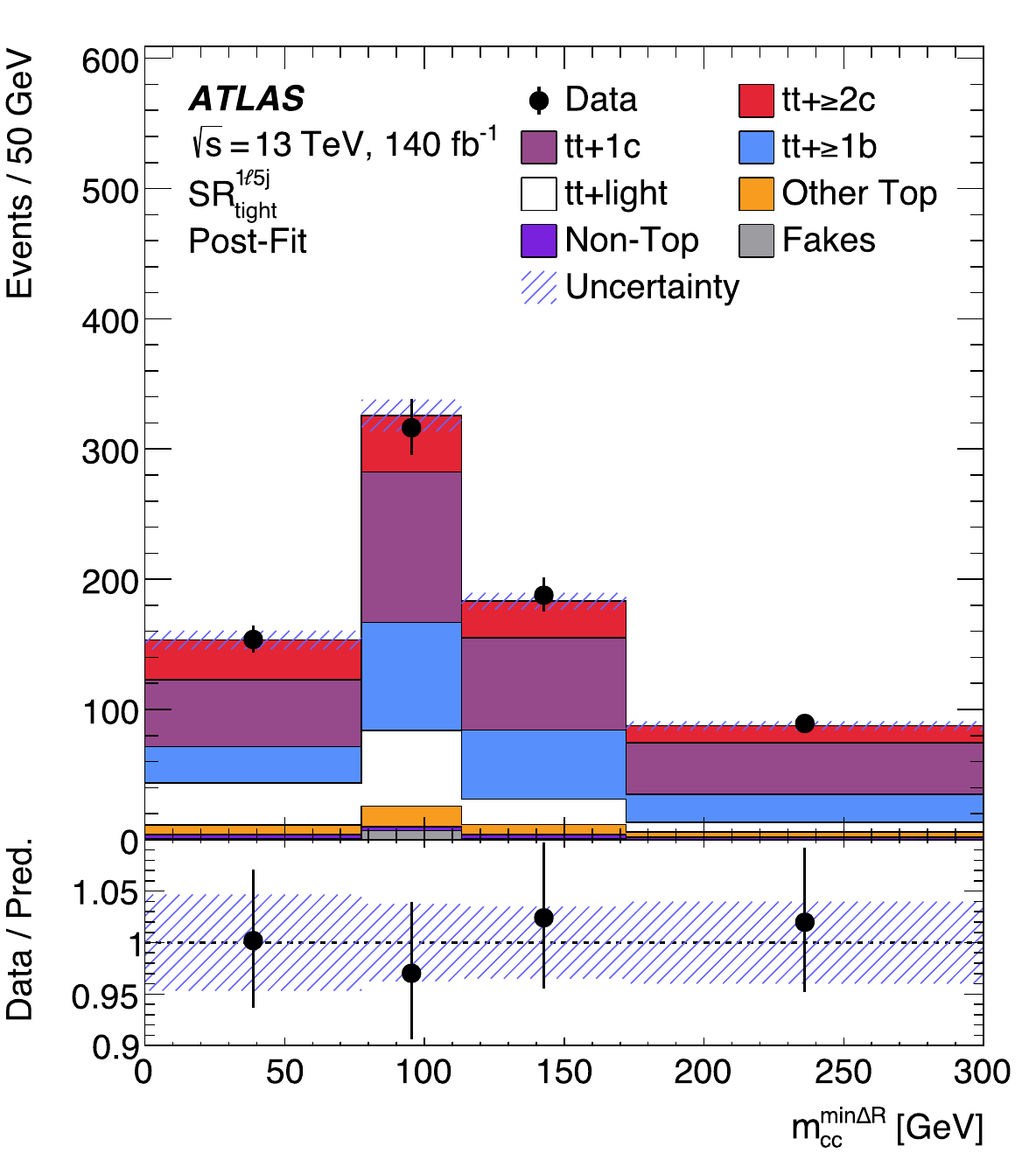}}\\
\subcaptionbox{\label{fig:post-fit-plot-3j-loose}}{\includegraphics[width=0.48\textwidth]{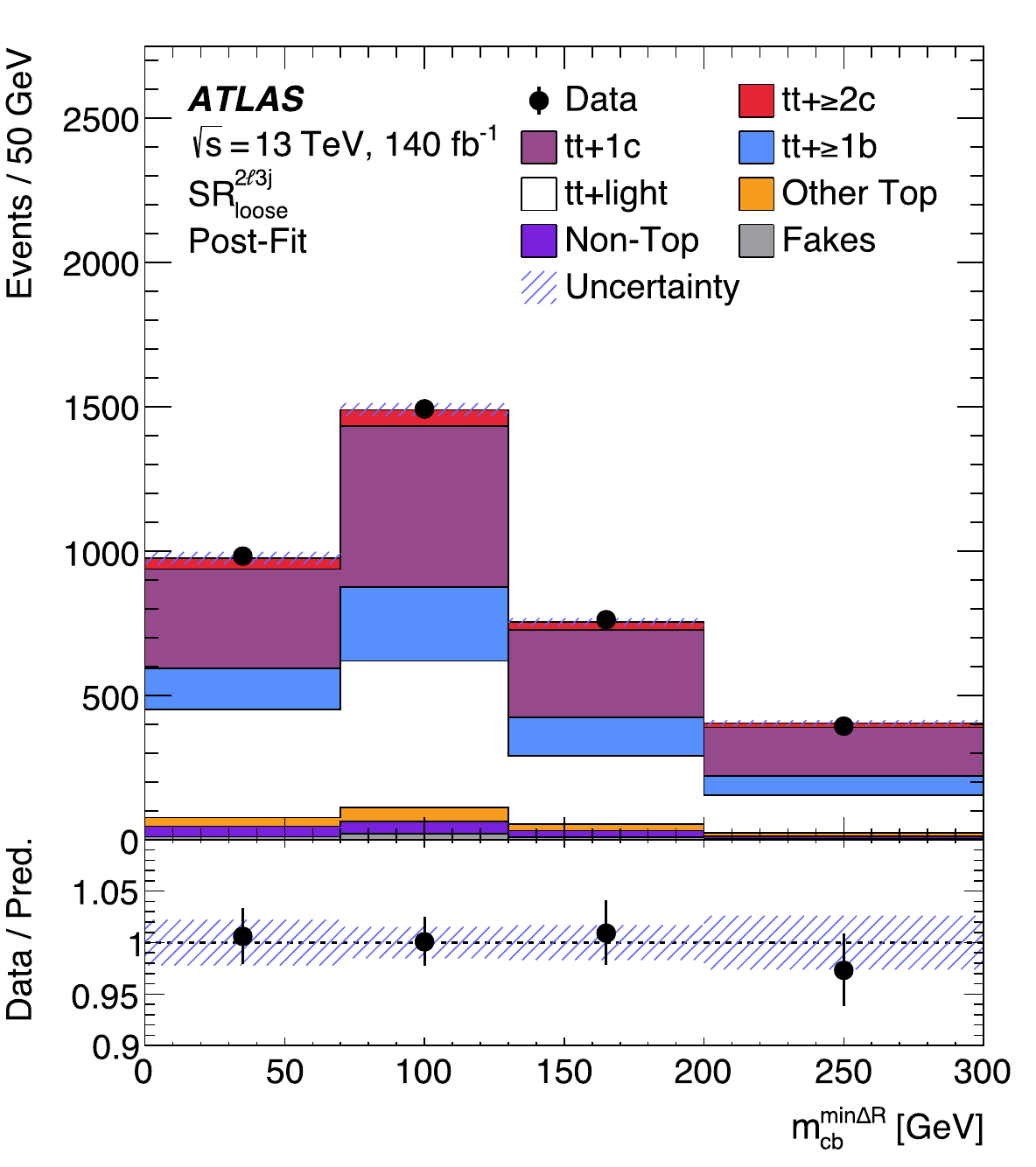}}
\caption{
Post-fit agreement between data and \ac{MC} prediction for the observables used in the 5-jet-exclusive and 3-jet-exclusive \acfp{SR}:
\protect\subref{fig:post-fit-plot-5j-loose} \srlooselfivej and \protect\subref{fig:post-fit-plot-5j-tight} \srtightlfivej, both using the invariant mass of the two geometrically closest \ctagged jets, $m_{\mathit{cc}}^{\mathrm{min}\Delta R}$, for the latter only considering \ctagged jets passing \bctightc;
\protect\subref{fig:post-fit-plot-3j-loose} \srloosellthreej using the invariant mass of the selected \ctagged jet and the geometrically closest \btagged jet, $m_{\mathit{cb}}^{\mathrm{min}\Delta R}$.
The hatched uncertainty bands include all uncertainties and their correlations.
The last bins contain overflow events.
\emph{Other Top} includes single-top-quark production and associated production of \ttbar and single top quarks with bosons.
\emph{Non-Top} includes \Wjets, \Zjets, and diboson processes.
}
\label{fig:post-fit-plots-3j-5j}
\end{figure}

\begin{figure}[htp]
\centering
\subcaptionbox{\label{fig:post-fit-plot-6ji-loose}}{\includegraphics[width=0.48\textwidth]{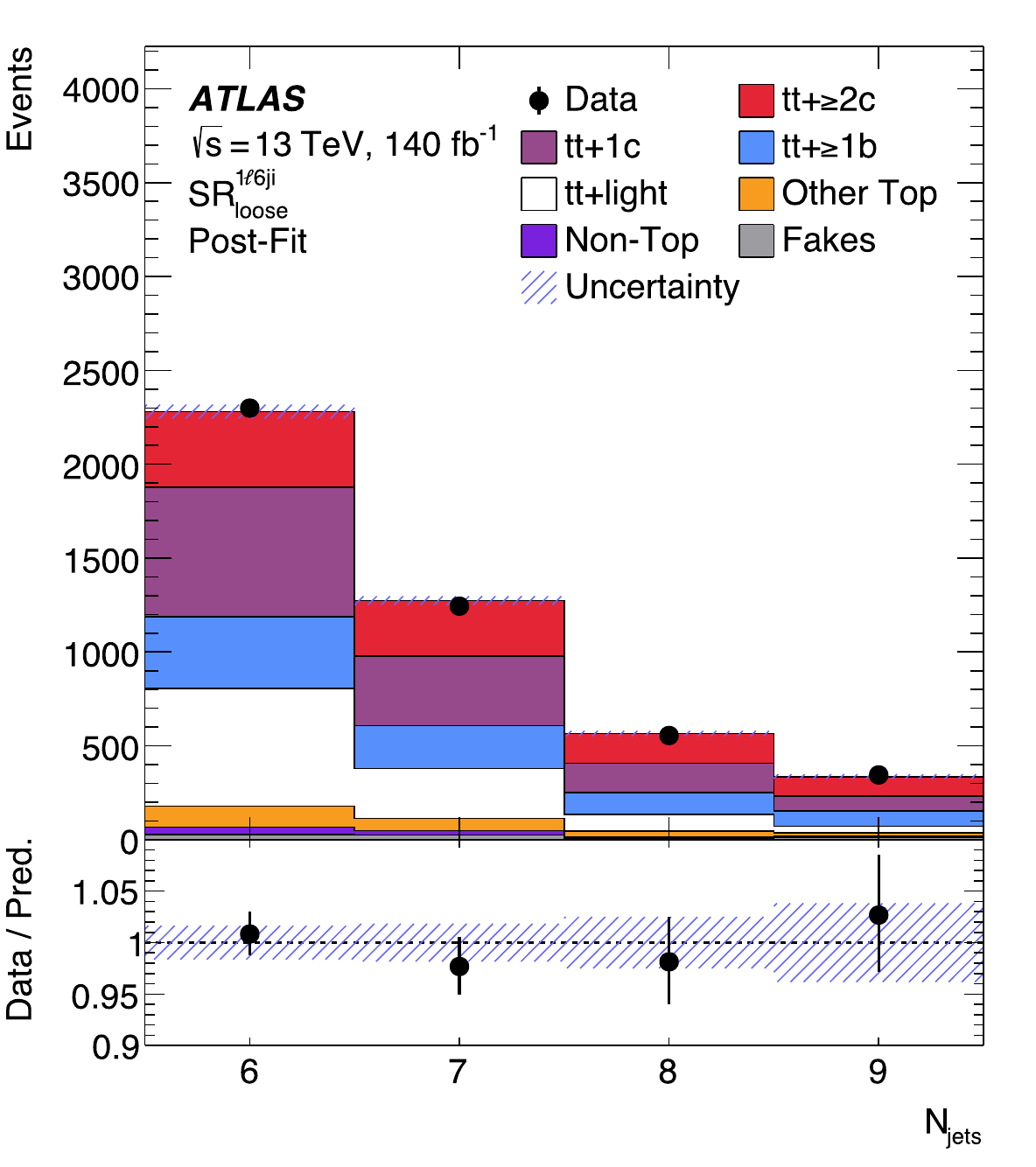}}\hfill
\subcaptionbox{\label{fig:post-fit-plot-6ji-tight}}{\includegraphics[width=0.48\textwidth]{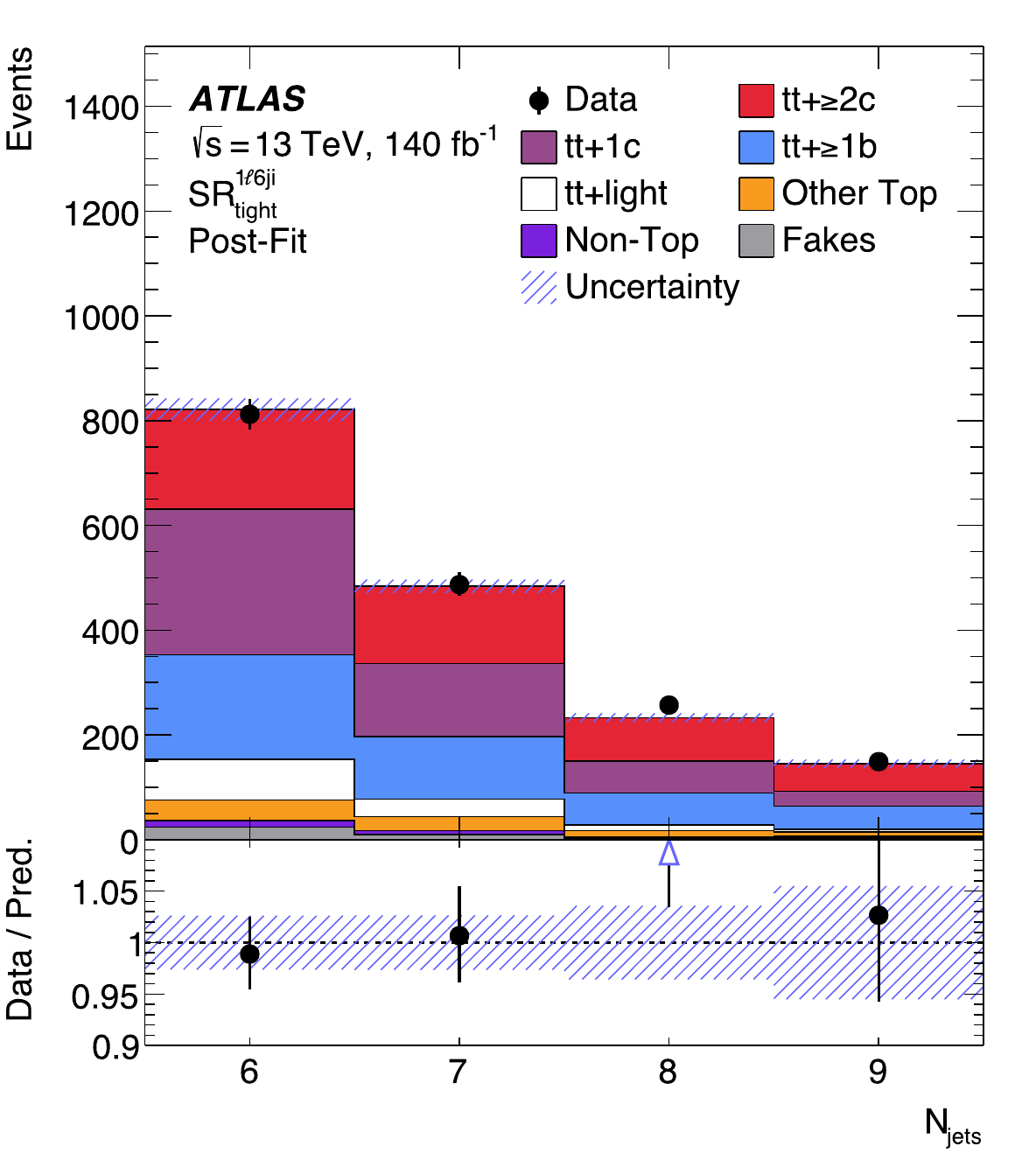}}\\
\subcaptionbox{\label{fig:post-fit-plot-4ji-loose}}{\includegraphics[width=0.48\textwidth]{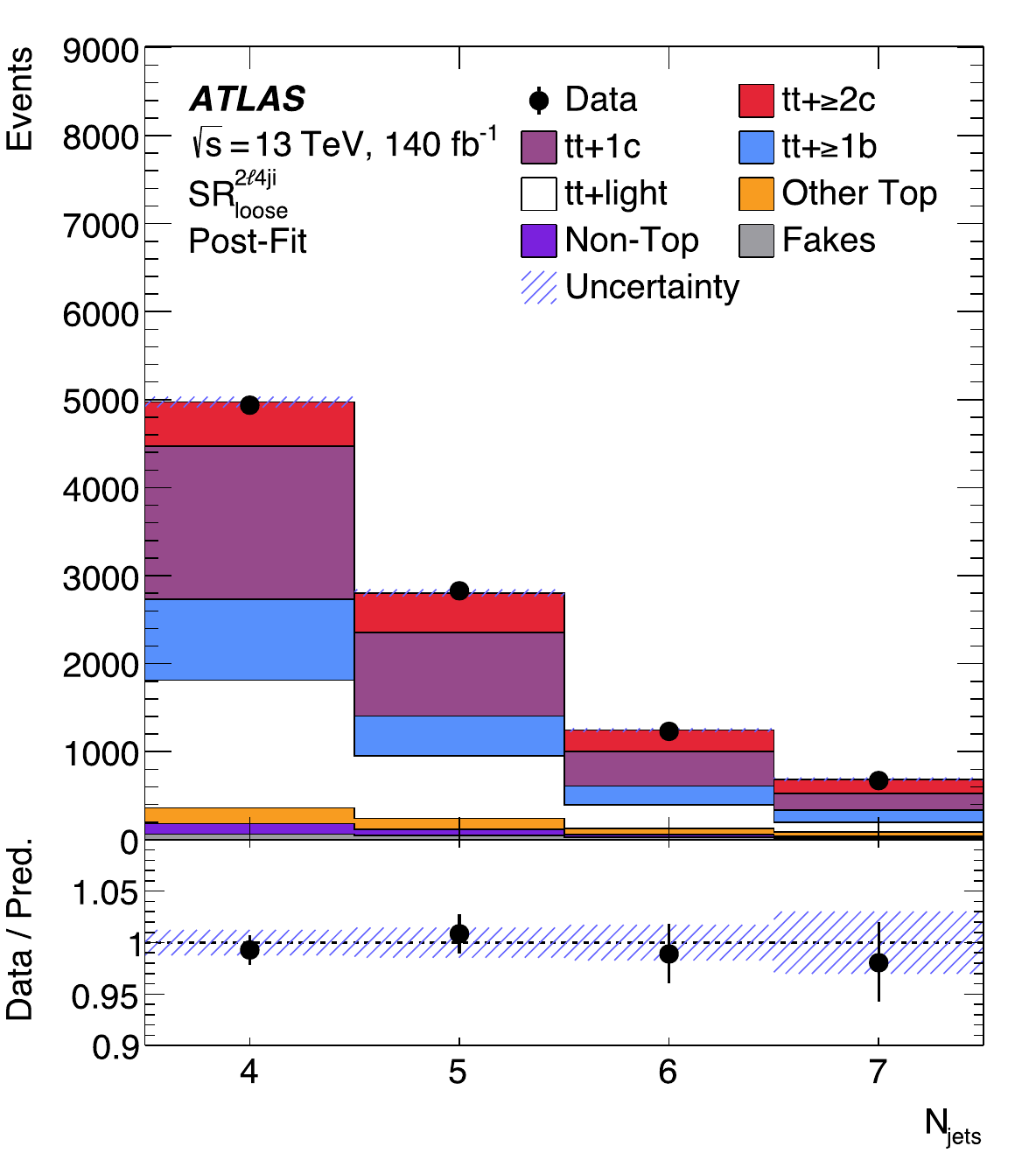}}\hfill
\subcaptionbox{\label{fig:post-fit-plot-4ji-tight}}{\includegraphics[width=0.48\textwidth]{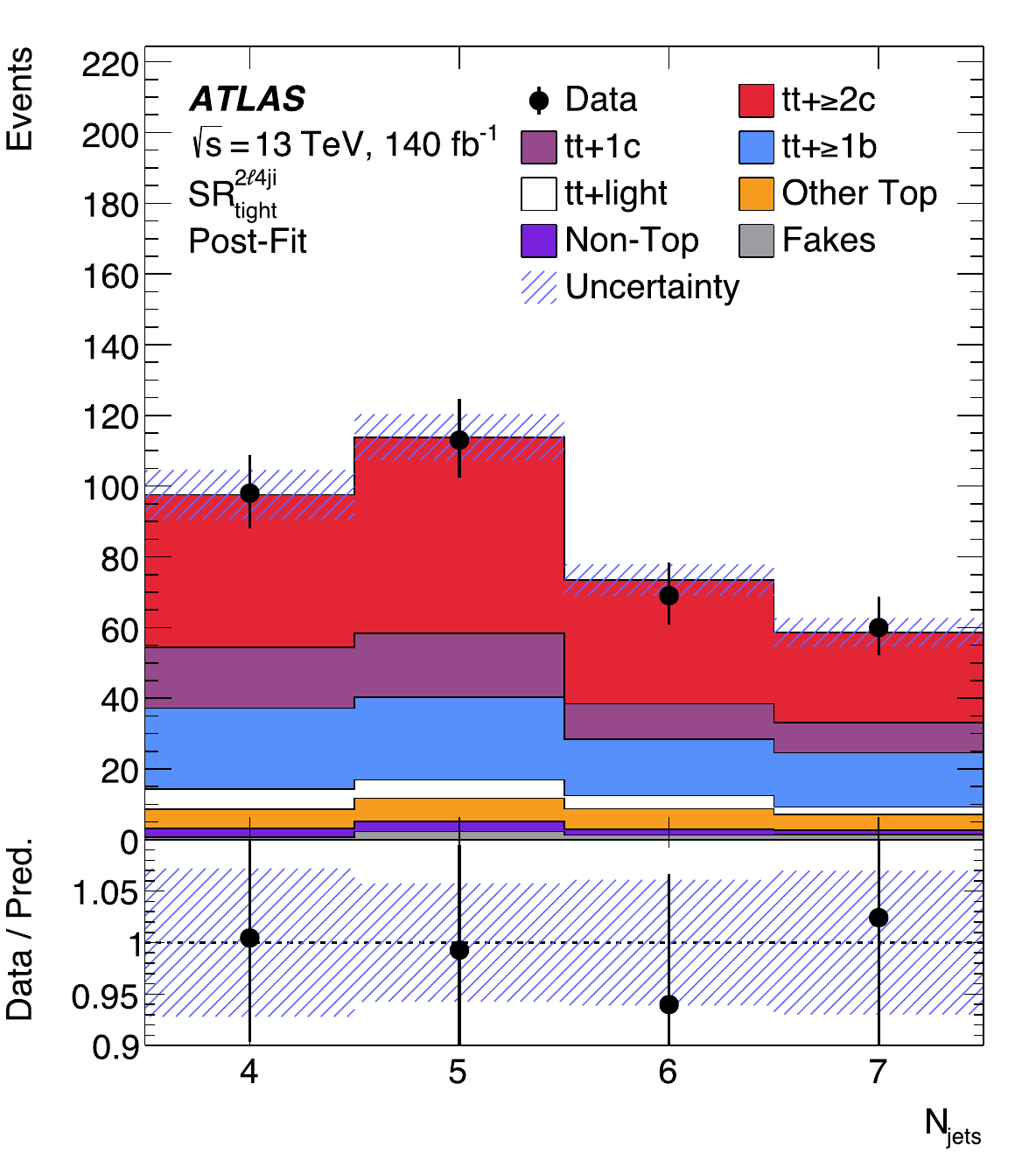}}
\caption{%
Post-fit agreement between data and \ac{MC} prediction for the observables used in the
\protect\subref{fig:post-fit-plot-6ji-loose} loose 6-jet-inclusive,
\protect\subref{fig:post-fit-plot-6ji-tight} tight 6-jet-inclusive,
\protect\subref{fig:post-fit-plot-4ji-loose} loose 4-jet-inclusive, and
\protect\subref{fig:post-fit-plot-4ji-tight} tight 4-jet-inclusive \acfp{SR}.
All regions use the jet multiplicity, $N_{\mathrm{jets}}$, as a fit observable.
The hatched uncertainty bands include all uncertainties and their correlations.
The last bins contain overflow events.
\emph{Other Top} includes single-top-quark production and associated production of \ttbar and single top quarks with bosons.
\emph{Non-Top} includes \Wjets, \Zjets, and diboson processes.
}
\label{fig:post-fit-plots-4ji-6ji}
\end{figure}

Post-fit comparisons between data and simulation for all regions are depicted in \cref{fig:post_fit_regions}.
The fitted observable distributions of the \acp{SR} are shown in \cref{fig:post-fit-plots-3j-5j,fig:post-fit-plots-4ji-6ji}.
Good agreement between data and simulation is observed in all regions and observables with varying relative contributions of the \ttjets categories.
The goodness of fit was evaluated using a \emph{saturated model}~\cite{BAKER1984437} and the compatibility with data was found to be \SI{98}{\percent}.

The \ttccin and \ttc normalization factors obtained from this fit are used to extract the observed fiducial \xsecs by scaling them with the predicted \xsecs for the fiducial phase space introduced in \cref{sec:simulation}.
Theoretical uncertainties in the predicted \xsecs owing to the \muR and \muF scale choice and the used \ac{PDF} set are not propagated to the extracted fiducial \xsec values.
The fiducial \xsecs for \ttccin and \ttc processes are determined to be
\begin{align}
\sigmattccin
& = \num{1.28} \, ^{+0.16}_{-0.10} \, \text{(stat)} \, ^{+0.21}_{-0.22} \, \text{(syst)} \; \unit{\pb}
= 1.28 \, ^{+0.27}_{-0.24} \; \unit{\pb},\\
\sigmattc
&= \num{6.4} \, ^{+0.5}_{-0.4} \, \text{(stat)} \, \pm 0.8 \, \text{(syst)} \; \unit{\pb}
= 6.4 \, ^{+1.0}_{-0.9} \; \unit{\pb}.
\end{align}
The breakdown into statistical and systematic uncertainties is estimated from the covariance matrix of the profile likelihood fit, following Ref.~\cite{Pinto:2023yob}.
The precision of the measured \ttccin and \ttc \xsecs is limited by uncertainties in the modeling of \ttcin, \ttbin, and \ttlight, in particular in the \ac{NLO} matching and the \ac{PS}, by uncertainties in the \bctagger calibration, as well as by data statistics.
\Cref{tab:grouped-impact} provides a detailed breakdown of the uncertainties that affect the \ttccin and \ttc \xsecs in the fiducial phase space.
The uncertainty in the \ttbin and \ttlight normalization factors on the \ttccin and \ttc \xsecs is included in the data statistics entry.
The fit constrains several nuisance parameters that strongly impact the \ttccin and \ttc \xsecs, with none falling below 50\% of their prior values.
Noteworthy constraints are placed on the \ttcin \ac{FSR} and \ttc \ac{PS} nuisance parameters, which arise primarily from discrepancies between the predicted rates of \ttccin and \ttc events in the signal regions.
No significant pulls of the nuisance parameters are observed in the fit.
The measured \ttbin normalization factor is compatible with those obtained in a dedicated ATLAS \ttbb measurement performed in a fiducial phase space targeting $e+\mu$ final states~\cite{TOPQ-2019-03}, reaching a similar level of relative precision.

\begin{figure}
\centering
\includegraphics[width=1.0\textwidth]{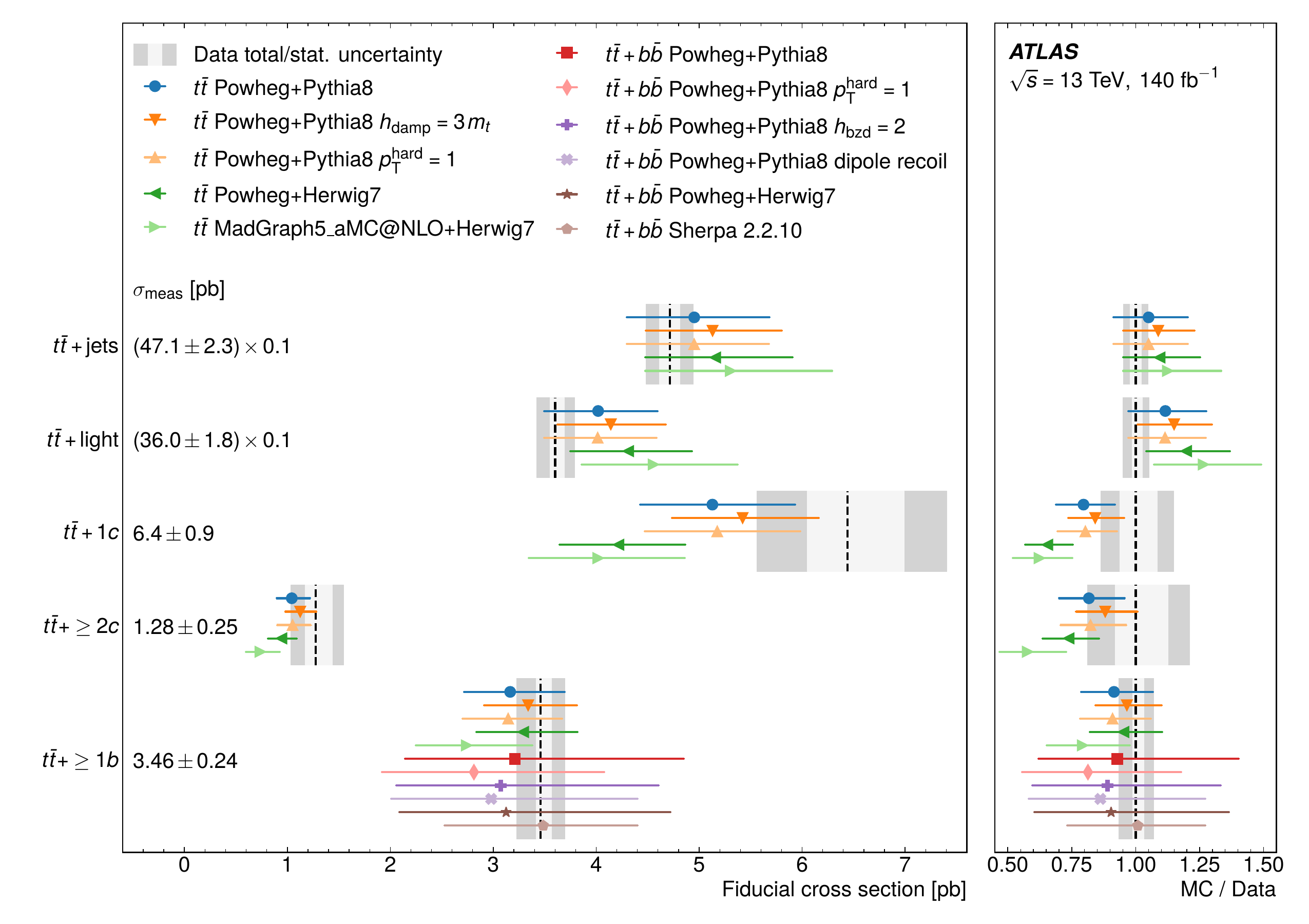}
\caption{
Measured fiducial cross-section values in comparison with various \ac{NLO+PS} predictions from inclusive \ttbar and \ttbb simulations.
The boxes in the background represent the statistical and total uncertainties of the values extracted from data.
The uncertainty bars of the predictions include independent and simultaneous variations of the \muR and \muF scales and uncertainties in the choice of the \ac{PDF} set, but no uncertainties in the parton shower or hadronization.
The measured and computed values for \ttjets and \ttlight are scaled by a factor of \num{0.1} to facilitate visualization.
For presentation purposes, the quoted measurement uncertainties are symmetrized.
}
\label{fig:fid_xsecs}
\end{figure}

The measured values for the four processes and the overall \ttjets production are depicted in \cref{fig:fid_xsecs} and listed in \cref{tab:fid_xsec_results}, along with several \ac{NLO+PS} predictions from different \ac{MC} simulations.
The predictions for \ttccin and \ttc are largely consistent with the measurements, but underpredict the observed \xsecs by 0.5 to 2.0 standard deviations.
Considering variations of the \muR and \muF scales and uncertainties in the \ac{PDF} choice of the predictions, the \POWHEG{}+\PYTHIA{8} setups agree with the measured \ttccin (\ttc) values within 0.5 to 0.8 (0.9 to 1.1) standard deviations of measurement and prediction uncertainties.
In contrast, \POWHEG{}+\HERWIG{7} and \MGNLO{}+\HERWIG{7} fall short by 25\% to 40\%, aligning with the measured values at 1.2 to 2.0 standard deviations.
For \ttbin, \ttlight, and \ttjets, most \ac{NLO+PS} predictions agree with the measured \xsecs within measurement uncertainties.

\begin{table}
\centering
\caption{%
Breakdown of the fiducial \ttccin and \ttc \xsec uncertainties.
Systematic uncertainties are grouped into signal and background modeling, instrumental, and \ac{MC} statistics categories.
The table lists the fractional uncertainty in the measured \xsecs in percent and is estimated from the covariance matrix of the profile likelihood fit, following Ref.~\cite{Pinto:2023yob}.
Individual groups of uncertainties can be added in quadrature to obtain the total uncertainty.
JES and JER denote the jet energy scale and resolution uncertainties, respectively.
The fractional uncertainties in the fitted \ttbin and \ttlight normalization factors are included in the data statistics category.
For presentation purposes, all uncertainties are symmetrized.
}
\label{tab:grouped-impact}
\sisetup{table-format=2.1}
\footnotesize
\begin{tabular}[pos]{l S S}
\toprule
Uncertainty group            & \multicolumn{2}{c}{Fractional uncertainty [\%] on} \\
& {\sigmattccin}   & {\sigmattc} \\
\midrule
\ttcin modeling              & 9   & 8   \\
Background modeling:         &     &     \\
\quad\quad\ttbin             & 4   & 4   \\
\quad\quad\ttlight           & 6   & 4   \\
\quad\quad Others            & 2.5 & 1.7 \\
Instrumental: & & \\
\quad\quad $b$-tagging       & 2.2 & 1.8 \\
\quad\quad $c$-tagging       & 9   & 4   \\
\quad\quad light mis-tagging & 2.2 & 3.4 \\
\quad\quad JES/JER           & 6   & 3.5 \\
\quad\quad Others            & 1.3 & 0.9 \\
MC statistics                & 3.1 & 2.5 \\
\midrule
\textbf{Total systematic uncertainty} & 17  & 12  \\
\midrule
Data statistical uncertainty & 11  & 7   \\
\midrule
\textbf{Total}               & 20  & 14  \\
\bottomrule
\end{tabular}
\end{table}

\begin{table}
\centering
\caption{
Measured fiducial \xsec values in comparison with various \ac{NLO+PS} predictions.
The uncertainties in the predictions include independent and simultaneous variations of the \muR and \muF scales and uncertainties in the choice of the \ac{PDF} set, but no uncertainties in the parton shower or hadronization.
The uncertainties in the measured values include all statistical and systematic uncertainties.
For presentation purposes, the quoted measurement and prediction uncertainties are symmetrized.
}
\label{tab:fid_xsec_results}
\sisetup{table-format=1.1(1),separate-uncertainty}
\footnotesize
\begin{tabular}{
l
S[table-format=1.2(2)]
S[table-format=1.1(1)]
S[table-format=1.1(1)]
S[table-format=2.0(1)]
S[table-format=2.0(1)]
}
\toprule
& {\ttccin[pb]} & {\ttc[pb]} & {\ttbin[pb]} & {\ttlight[pb]} & {\ttjets[pb]} \\
\midrule
$\ttbar$ \POWHEG{}+\PYTHIA[8]                  & 1.04 \pm 0.18  & 5.1 \pm 0.8 & 3.2 \pm 0.5   & 40 \pm 6     & 50 \pm 7     \\
$\ttbar$ \POWHEG{}+\PYTHIA[8], $\hdamp=3\,m_t$ & 1.12 \pm 0.16  & 5.4 \pm 0.7 & 3.3 \pm 0.5   & 41 \pm 5     & 51 \pm 7     \\
$\ttbar$ \POWHEG{}+\PYTHIA[8], $\pthard=1$     & 1.05 \pm 0.18  & 5.2 \pm 0.8 & 3.1 \pm 0.5   & 40 \pm 6     & 50 \pm 7     \\
$\ttbar$ \POWHEG{}+\HERWIG[7]                  & 0.94 \pm 0.16  & 4.2 \pm 0.7 & 3.3 \pm 0.5   & 43 \pm 6     & 52 \pm 8     \\
$\ttbar$ \MGNLO{}+\HERWIG[7]                   & 0.74 \pm 0.19  & 4.0 \pm 0.8 & 2.7 \pm 0.6   & 46 \pm 8     & 53 \pm 10    \\
\midrule
$\ttbb$ \POWHEG{}+\PYTHIA[8]                   & {---}          & {---}       & 3.2 \pm 1.6   & {---}        & {---}        \\
$\ttbb$ \POWHEG{}+\PYTHIA[8], $\pthard=1$      & {---}          & {---}       & 2.8 \pm 1.3   & {---}        & {---}        \\
$\ttbb$ \POWHEG{}+\PYTHIA[8], $\hbzd=2$        & {---}          & {---}       & 3.1 \pm 1.5   & {---}        & {---}        \\
$\ttbb$ \POWHEG{}+\PYTHIA[8], dipole recoil    & {---}          & {---}       & 3.0 \pm 1.4   & {---}        & {---}        \\
$\ttbb$ \POWHEG{}+\HERWIG[7]                   & {---}          & {---}       & 3.1 \pm 1.6   & {---}        & {---}        \\
$\ttbb$ \SHERPA[2.2.10]                        & {---}          & {---}       & 3.5 \pm 1.0   & {---}        & {---}        \\
\midrule
Data                                           & 1.28 \pm 0.25  & 6.4 \pm 0.9 & {$3.46 \pm 0.24$} & {$36.0 \pm 1.8$} & {$47.1 \pm 2.3$} \\
\bottomrule
\end{tabular}
\end{table}

To test the compatibility of the two channels, an additional fit was performed where the \ttccin and \ttc normalization factors were parameterized independently in the \slepton and \dilepton regions.
The \slepton channel measures $\sigmattccin = \SI{1.5 \pm 0.4}{\pb}$ and $\sigmattc = \SI{6.4 \pm 1.1}{\pb}$, and the \dilepton channel $\sigmattccin = \SI{1.18 \pm 0.25}{\pb}$ and $\sigmattc = \SI{6.4 \pm 1.0}{\pb}$.
The results in both channels are consistent with the nominal results within their uncertainties.
The compatibility of this fit with the nominal setup was evaluated by performing a $\chi^2$ test on the log-likelihood difference, yielding a compatibility of \SI{58}{\percent}.
In a second test, the \ttccin and \ttc normalization factors were parameterized through one joint parameter of interest, and the \ac{NLO} matching and \ac{PS} uncertainties of the two processes were treated as correlated.
The resulting \xsec of $\sigma^{\mathrm{fid}}(\ttcin) = \SI{8.2 \pm 0.9}{\pb}$ is consistent with the sum of the \ttccin and \ttc \xsecs in the nominal setup within uncertainties, but shows increased relative precision.

A measurement of the \xsecs in the more inclusive phase space yields $\sigmattccinfull = \SI{5.4 \pm 1.1}{\pb}$ and $\sigmattcfull = \SI{38 \pm 6}{\pb}$, showing moderately increased relative uncertainties compared with the fiducial \xsecs.

The normalization factors \muttbin, \muttccin, \muttc, and \muttlight are used to determine the \xsec ratios of these processes to total \ttjets production.
These measurements benefit from the cancelation of several systematic uncertainties related to detector instrumentation and calibrations, potentially reducing the overall systematic uncertainties.
In the more inclusive phase space, the measurement yields ratios of $\Rttccininc = \SI{1.23 \pm 0.25}{\percent}$ and $\Rttcinc = \SI{8.8 \pm 1.3}{\percent}$.
In the fiducial phase space, these ratios increase to $\Rttccin = \SI{2.7 \pm 0.5}{\percent}$ and $\Rttc = \SI{13.7 \pm 1.8}{\percent}$.
In both phase spaces, the \POWHEG{}+\PYTHIA{8} simulations underpredict the ratios, but agree with the measured \Rttccininc and \Rttcinc values within 0.9 and 1.1 standard deviations, and in the fiducial phase space within 1.0 and 1.4 standard deviations, respectively.
A~summary of all results is given in \cref{tab:results-summary}.

\begin{table}[tp]
\centering
\caption{%
Measured and predicted values for the \ttbin, \ttccin, and \ttc \xsecs and for the \xsec ratios of these processes to total \ttjets production.
The quoted uncertainties in the measurements include statistical and systematic uncertainties.
The predictions are taken from the nominal setup using
inclusive \ttbar \POWHEG{}+\PYTHIA{8} predictions for the \ttccin, \ttc, and \ttlight processes, and \ttbb \POWHEG{}+\PYTHIA{8} predictions for the \ttbin process.
The first use the \acf{5FS}, the latter the \acf{4FS}, where the $b$-quarks are treated as massive particles and are included in the matrix element calculation.
The uncertainties in the predictions include independent and simultaneous variations of the \muR and \muF scales and uncertainties in the choice of the \ac{PDF} set.
}
\label{tab:results-summary}
\footnotesize
\begin{tabular}{lr@{$\,\pm\,$}l r@{$\,\pm\,$}l}
\toprule
& \multicolumn{2}{c}{\multirow{2}{*}{Measured}} & \multicolumn{2}{c}{\ttbar or \ttbb} \\
& \multicolumn{2}{c}{} & \multicolumn{2}{c}{\POWHEG{}+\PYTHIA{8}} \\
\midrule
\sigmattbin[pb]      & 3.46 & 0.24  & \hspace{1.5em} 3.2  & 1.6  \\
\sigmattccin[pb]     & 1.28 & 0.25  & 1.04 & 0.18 \\
\sigmattc[pb]        & 6.4  & 0.9   & 5.1  & 0.8  \\
\midrule
\sigmattbinfull[pb]  & 13.0 & 0.9 & 12  & 4   \\
\sigmattccinfull[pb] &  5.4 & 1.1 & 4.4 & 0.7 \\
\sigmattcfull[pb]    & 38   & 6   & 31  & 4   \\
\midrule
\Rttbin[\%]          &  7.2 & 0.4   &  6.5 & 3.3  \\
\Rttccin[\%]         &  2.7 & 0.5   &  2.1 & 0.4  \\
\Rttc[\%]            & 13.7 & 1.8   & 10.3 & 1.6  \\
\midrule
\Rttbinfull[\%]      & 3.14 & 0.23  & 2.6  & 0.8  \\
\Rttccininc[\%]     & 1.23 & 0.25  & 0.97 & 0.16 \\
\Rttcinc[\%]        & 8.8  & 1.3   & 6.9  & 1.0  \\
\bottomrule
\end{tabular}
\end{table}

\FloatBarrier

\section{Conclusion}

The production of \ttbar pairs in association with heavy-flavor jets, \ttbb and \ttcc, is a background in many measurements of rare \ac{SM} processes and searches for new physics, yet a challenging process to model precisely.
This Letter presents a measurement of the production of \ttbar with additional jets initiated by charm quarks with the ATLAS experiment at the \ac{LHC}.
The analysis exploits the full ATLAS Run~2 \pp collision data sample at a center-of-mass energy of \SI{13}{\TeV}, corresponding to an integrated luminosity of \SI{140}{\ifb}, and considers events with one or two charged leptons in the final state.
A custom flavor tagging algorithm, termed the \bctagger, tailored to simultaneously tag \cjets and \bjets, is employed to define analysis regions sensitive to both \ttccin and \ttc production which are determined separately for the first time.
Using a profile likelihood fit approach, the \xsecs for \ttbar production with two or more additional \cjets and one additional \cjet are found to be $\sigmattccin = 1.28^{+0.27}_{-0.24}\;\unit{\pb}$ and $\sigmattc = 6.4^{+1.0}_{-0.9}\;\unit{\pb}$ in a fiducial volume that mimics the acceptance of the ATLAS detector.
\ac{NLO+PS} predictions for \ttccin and \ttc are largely consistent with the measurements, though all underpredict the observed values.
\POWHEG{}+\PYTHIA{8} setups agree with the measured values within 0.5 to 1.1 standard deviations, while \POWHEG{}+\HERWIG{7} and \MGNLO{}+\HERWIG{7} predict \xsecs up to 40\% lower than the measurements, with agreement observed at the level of up to 2.0 standard deviations.
The precision of the measured \xsecs is limited by uncertainties in the modeling of \ttcin, \ttbin, and \ttlight, and in the \bctagger calibration, as well as by data statistics.
The \xsec ratios of \ttccin and \ttc to total \ttjets production are found to be $\Rttccininc = \SI{1.23 \pm 0.25}{\percent}$ and $\Rttcinc = \SI{8.8 \pm 1.3}{\percent}$ in a phase-space volume without requirements on the \ttbar decay products and the jet multiplicity.
These results provide crucial inputs for improving the modeling of \ttbar production with additional \cjets, which would benefit from \ac{3FS} \ttcc simulations that incorporate the gluon splitting to a \ccbar pair into the matrix element calculation.
Understanding \ttcin production is essential for precision measurements of even rarer \ac{SM} processes and for searches for physics beyond the \ac{SM}, where \ttcin is an important background.

\section*{Acknowledgements}
%

%
%

%
%

We thank CERN for the very successful operation of the LHC and its injectors, as well as the support staff at
CERN and at our institutions worldwide without whom ATLAS could not be operated efficiently.

The crucial computing support from all WLCG partners is acknowledged gratefully, in particular from CERN, the ATLAS Tier-1 facilities at TRIUMF/SFU (Canada), NDGF (Denmark, Norway, Sweden), CC-IN2P3 (France), KIT/GridKA (Germany), INFN-CNAF (Italy), NL-T1 (Netherlands), PIC (Spain), RAL (UK) and BNL (USA), the Tier-2 facilities worldwide and large non-WLCG resource providers. Major contributors of computing resources are listed in Ref.~\cite{ATL-SOFT-PUB-2023-001}.

We gratefully acknowledge the support of ANPCyT, Argentina; YerPhI, Armenia; ARC, Australia; BMWFW and FWF, Austria; ANAS, Azerbaijan; CNPq and FAPESP, Brazil; NSERC, NRC and CFI, Canada; CERN; ANID, Chile; CAS, MOST and NSFC, China; Minciencias, Colombia; MEYS CR, Czech Republic; DNRF and DNSRC, Denmark; IN2P3-CNRS and CEA-DRF/IRFU, France; SRNSFG, Georgia; BMBF, HGF and MPG, Germany; GSRI, Greece; RGC and Hong Kong SAR, China; ISF and Benoziyo Center, Israel; INFN, Italy; MEXT and JSPS, Japan; CNRST, Morocco; NWO, Netherlands; RCN, Norway; MNiSW, Poland; FCT, Portugal; MNE/IFA, Romania; MSTDI, Serbia; MSSR, Slovakia; ARIS and MVZI, Slovenia; DSI/NRF, South Africa; MICIU/AEI, Spain; SRC and Wallenberg Foundation, Sweden; SERI, SNSF and Cantons of Bern and Geneva, Switzerland; NSTC, Taipei; TENMAK, T\"urkiye; STFC/UKRI, United Kingdom; DOE and NSF, United States of America.

Individual groups and members have received support from BCKDF, CANARIE, CRC and DRAC, Canada; CERN-CZ, FORTE and PRIMUS, Czech Republic; COST, ERC, ERDF, Horizon 2020, ICSC-NextGenerationEU and Marie Sk{\l}odowska-Curie Actions, European Union; Investissements d'Avenir Labex, Investissements d'Avenir Idex and ANR, France; DFG and AvH Foundation, Germany; Herakleitos, Thales and Aristeia programmes co-financed by EU-ESF and the Greek NSRF, Greece; BSF-NSF and MINERVA, Israel; NCN and NAWA, Poland; La Caixa Banking Foundation, CERCA Programme Generalitat de Catalunya and PROMETEO and GenT Programmes Generalitat Valenciana, Spain; G\"{o}ran Gustafssons Stiftelse, Sweden; The Royal Society and Leverhulme Trust, United Kingdom.

In addition, individual members wish to acknowledge support from Armenia: Yerevan Physics Institute (FAPERJ); CERN: European Organization for Nuclear Research (CERN PJAS); Chile: Agencia Nacional de Investigaci\'on y Desarrollo (FONDECYT 1230812, FONDECYT 1230987, FONDECYT 1240864); China: Chinese Ministry of Science and Technology (MOST-2023YFA1605700), National Natural Science Foundation of China (NSFC - 12175119, NSFC 12275265, NSFC-12075060); Czech Republic: Czech Science Foundation (GACR - 24-11373S), Ministry of Education Youth and Sports (FORTE CZ.02.01.01/00/22\_008/0004632), PRIMUS Research Programme (PRIMUS/21/SCI/017); EU: H2020 European Research Council (ERC - 101002463); European Union: European Research Council (ERC - 948254, ERC 101089007), European Union, Future Artificial Intelligence Research (FAIR-NextGenerationEU PE00000013), Italian Center for High Performance Computing, Big Data and Quantum Computing (ICSC, NextGenerationEU); France: Agence Nationale de la Recherche (ANR-20-CE31-0013, ANR-21-CE31-0013, ANR-21-CE31-0022, ANR-22-EDIR-0002); Germany: Baden-Württemberg Stiftung (BW Stiftung-Postdoc Eliteprogramme), Deutsche Forschungsgemeinschaft (DFG - 469666862, DFG - CR 312/5-2); Italy: Istituto Nazionale di Fisica Nucleare (ICSC, NextGenerationEU), Ministero dell'Università e della Ricerca (PRIN - 20223N7F8K - PNRR M4.C2.1.1); Japan: Japan Society for the Promotion of Science (JSPS KAKENHI JP22H01227, JSPS KAKENHI JP22H04944, JSPS KAKENHI JP22KK0227, JSPS KAKENHI JP23KK0245); Netherlands: Netherlands Organisation for Scientific Research (NWO Veni 2020 - VI.Veni.202.179); Norway: Research Council of Norway (RCN-314472); Poland: Ministry of Science and Higher Education (IDUB AGH, POB8, D4 no 9722), Polish National Agency for Academic Exchange (PPN/PPO/2020/1/00002/U/00001), Polish National Science Centre (NCN 2021/42/E/ST2/00350, NCN OPUS 2023/51/B/ST2/02507, NCN OPUS nr 2022/47/B/ST2/03059, NCN UMO-2019/34/E/ST2/00393, NCN \& H2020 MSCA 945339, UMO-2020/37/B/ST2/01043, UMO-2021/40/C/ST2/00187, UMO-2022/47/O/ST2/00148, UMO-2023/49/B/ST2/04085, UMO-2023/51/B/ST2/00920); Slovenia: Slovenian Research Agency (ARIS grant J1-3010); Spain: Generalitat Valenciana (Artemisa, FEDER, IDIFEDER/2018/048), Ministry of Science and Innovation (MCIN \& NextGenEU PCI2022-135018-2, MICIN \& FEDER PID2021-125273NB, RYC2019-028510-I, RYC2020-030254-I, RYC2021-031273-I, RYC2022-038164-I); Sweden: Carl Trygger Foundation (Carl Trygger Foundation CTS 22:2312), Swedish Research Council (Swedish Research Council 2023-04654, VR 2018-00482, VR 2022-03845, VR 2022-04683, VR 2023-03403, VR grant 2021-03651), Knut and Alice Wallenberg Foundation (KAW 2018.0458, KAW 2019.0447, KAW 2022.0358); Switzerland: Swiss National Science Foundation (SNSF - PCEFP2\_194658); United Kingdom: Leverhulme Trust (Leverhulme Trust RPG-2020-004), Royal Society (NIF-R1-231091); United States of America: U.S. Department of Energy (ECA DE-AC02-76SF00515), Neubauer Family Foundation.

%
%


%
\printbibliography
\clearpage
 
\begin{flushleft}
\hypersetup{urlcolor=black}
{\Large The ATLAS Collaboration}

\bigskip

\AtlasOrcid[0000-0002-6665-4934]{G.~Aad}$^\textrm{\scriptsize 105}$,
\AtlasOrcid[0000-0001-7616-1554]{E.~Aakvaag}$^\textrm{\scriptsize 17}$,
\AtlasOrcid[0000-0002-5888-2734]{B.~Abbott}$^\textrm{\scriptsize 124}$,
\AtlasOrcid[0000-0002-0287-5869]{S.~Abdelhameed}$^\textrm{\scriptsize 120a}$,
\AtlasOrcid[0000-0002-1002-1652]{K.~Abeling}$^\textrm{\scriptsize 57}$,
\AtlasOrcid[0000-0001-5763-2760]{N.J.~Abicht}$^\textrm{\scriptsize 51}$,
\AtlasOrcid[0000-0002-8496-9294]{S.H.~Abidi}$^\textrm{\scriptsize 30}$,
\AtlasOrcid[0009-0003-6578-220X]{M.~Aboelela}$^\textrm{\scriptsize 46}$,
\AtlasOrcid[0000-0002-9987-2292]{A.~Aboulhorma}$^\textrm{\scriptsize 36e}$,
\AtlasOrcid[0000-0001-5329-6640]{H.~Abramowicz}$^\textrm{\scriptsize 156}$
% ...
\bigskip
\\

% ...
$^{17}$Department for Physics and Technology, University of Bergen, Bergen; Norway.\\
% ...
$^{30}$Physics Department, Brookhaven National Laboratory, Upton NY; United States of America.\\
% ...
$^{36}$$^{(a)}$Facult\'e des Sciences Ain Chock, Universit\'e Hassan II de Casablanca;$^{(b)}$Facult\'{e} des Sciences, Universit\'{e} Ibn-Tofail, K\'{e}nitra;$^{(c)}$Facult\'e des Sciences Semlalia, Universit\'e Cadi Ayyad, LPHEA-Marrakech;$^{(d)}$LPMR, Facult\'e des Sciences, Universit\'e Mohamed Premier, Oujda;$^{(e)}$Facult\'e des sciences, Universit\'e Mohammed V, Rabat;$^{(f)}$Institute of Applied Physics, Mohammed VI Polytechnic University, Ben Guerir; Morocco.\\
% ...
$^{46}$Physics Department, Southern Methodist University, Dallas TX; United States of America.\\
% ...
$^{51}$Fakult\"{a}t Physik , Technische Universit{\"a}t Dortmund, Dortmund; Germany.\\
% ...
$^{57}$II. Physikalisches Institut, Georg-August-Universit\"{a}t G\"ottingen, G\"ottingen; Germany.\\
% ...
$^{105}$CPPM, Aix-Marseille Universit\'e, CNRS/IN2P3, Marseille; France.\\
% ...
$^{120}$$^{(a)}$New York University Abu Dhabi, Abu Dhabi;$^{(b)}$United Arab Emirates University, Al Ain; United Arab Emirates.\\
% ...
$^{124}$Homer L. Dodge Department of Physics and Astronomy, University of Oklahoma, Norman OK; United States of America.\\
% ...
$^{156}$Raymond and Beverly Sackler School of Physics and Astronomy, Tel Aviv University, Tel Aviv; Israel.\\
% ...
$^{a}$ Also Affiliated with an institute covered by a cooperation agreement with CERN.\\
$^{b}$ Also at An-Najah National University, Nablus; Palestine.\\
$^{c}$ Also at Borough of Manhattan Community College, City University of New York, New York NY; United States of America.\\
$^{d}$ Also at Center for Interdisciplinary Research and Innovation (CIRI-AUTH), Thessaloniki; Greece.\\
$^{e}$ Also at CERN, Geneva; Switzerland.\\
$^{f}$ Also at CMD-AC UNEC Research Center, Azerbaijan State University of Economics (UNEC); Azerbaijan.\\
% ...

\end{flushleft}


%

%
%

%

%
%
%
%

\end{document}